\documentclass[trackchanges, twocolumn, twocolappendix, resetfootnote]{aastex701}

\usepackage{graphicx}	
\usepackage{amsmath}	
\usepackage{graphicx}
\usepackage{subcaption}
\usepackage{array}
\usepackage{booktabs}
\usepackage{siunitx}
\usepackage{mathrsfs}
\usepackage{rotating}
\usepackage{tabularx}
\usepackage{longtable}
\usepackage{booktabs} 
\usepackage{afterpage}
\usepackage{xcolor}

\begin{document}

\title{Demonstrating the Time-Domain Capabilities of the 4-m International Liquid Mirror Telescope: An Early Census of Transient and Variable Detections}

\author[orcid=0000-0000-0000-0001,gname=Kumar,sname=Pranshu]{Kumar Pranshu}
\affiliation{Aryabhatta Research Institute of Observational Sciences (ARIES), Manora Peak, Nainital-263001, India}
\affiliation{Department of Applied Optics and Photonics, University of Calcutta, Kolkata, 700106, India}
\email[show]{kumarpranshu86@gmail.com}

\author[orcid=0000-0003-1637-267X,gname=Kuntal,sname=Misra]{Kuntal Misra}
\affiliation{Aryabhatta Research Institute of Observational Sciences (ARIES), Manora Peak, Nainital-263001, India}
\email{kuntal@aries.res.in}

\author[orcid=0009-0000-1020-9711,gname=Bhavya,sname=Ailawadhi]{Bhavya Ailawadhi}
\affiliation{Physical Research Laboratory, Navrangpura, Ahmedabad, Gujarat-380009, India}
\affiliation{Aryabhatta Research Institute of Observational Sciences (ARIES), Manora Peak, Nainital-263001, India}
\email{bhavya@prl.res.in}

\author[orcid=0009-0002-2621-6611,gname=Monalisa,sname=Dubey]{Monalisa Dubey}
\affiliation{Aryabhatta Research Institute of Observational Sciences (ARIES), Manora Peak, Nainital-263001, India}
\affiliation{Mahatma Jyotiba Phule Rohilkhand University, Pilibhit Bypass Road, Bareilly 243006, Uttar Pradesh, India}
\email{monalisa@aries.res.in}

\author[orcid=0000-0002-0394-6745,gname=Naveen,sname=Dukiya]{Naveen Dukiya}
\affiliation{Aryabhatta Research Institute of Observational Sciences (ARIES), Manora Peak, Nainital-263001, India}
\affiliation{Mahatma Jyotiba Phule Rohilkhand University, Pilibhit Bypass Road, Bareilly 243006, Uttar Pradesh, India}
\email{ndukiya@aries.res.in}

\author[orcid=0009-0009-3108-3789,gname=Sara,sname=Filali]{Sara Filali}
\affiliation{Institute of Astrophysics and Geophysics, Li\`ege University, All\'ee du 6 Ao\^ut 19c, 4000 Li\`ege, Belgium}
\email{sfilali.igi@gmail.com}

\author[gname=Paul,sname=Hickson]{Paul Hickson}
\affiliation{Department of Physics and Astronomy, The University of British Columbia, 6224 Agricultural Road, Vancouver, BC V6T 1Z1, Canada}
\affiliation{Outer Space Institute, The University of British Columbia, 325-6224 Agricultural Road, Vancouver, BC V6T 1Z1, Canada}
\email{hickson@physics.ubc.ca}

\author[gname=Priyanshi,sname=Kumari]{Priyanshi Kumari}
\affiliation{Aryabhatta Research Institute of Observational Sciences (ARIES), Manora Peak, Nainital-263001, India}
\email{kumaripriyanshi98@gmail.com}

\author[orcid=0009-0002-9445-3731,gname=Gokul Singh,sname=Mehra]{Gokul Singh Mehra}
\affiliation{Aryabhatta Research Institute of Observational Sciences (ARIES), Manora Peak, Nainital-263001, India}
\email{gokulmehra51@gmail.com}

\author[orcid=0000-0001-5824-1040,gname=Vibhore,sname=Negi]{Vibhore Negi}
\affiliation{Kavli Institute for Astronomy and Astrophysics, Peking University, Beijing 100871, China}
\affiliation{Aryabhatta Research Institute of Observational Sciences (ARIES), Manora Peak, Nainital-263001, India}
\email{vibhore.negi18@gmail.com}

\author[orcid=0000-0002-4331-1867,gname=Jeewan,sname=Pandey]{Jeewan C. Pandey}
\affiliation{Aryabhatta Research Institute of Observational Sciences (ARIES), Manora Peak, Nainital-263001, India}
\email{jeewan@aries.res.in}

\author[gname=Anna,sname=Pospieszalska-Surdej]{Anna Pospieszalska-Surdej}
\affiliation{Institute of Astrophysics and Geophysics, Li\`ege University, All\'ee du 6 Ao\^ut 19c, 4000 Li\`ege, Belgium}
\email{annapospie@gmail.com}

\author[orcid=0000-0002-7005-1976,gname=Jean,sname=Surdej]{Jean Surdej}
\affiliation{Institute of Astrophysics and Geophysics, Li\`ege University, All\'ee du 6 Ao\^ut 19c, 4000 Li\`ege, Belgium}
\affiliation{Aryabhatta Research Institute of Observational Sciences (ARIES), Manora Peak, Nainital-263001, India}
\email{jsurdej@uliege.be}

\author[orcid=0009-0002-2355-5626,gname=Sarvesh Kumar,sname=Yadav]{Sarvesh Kumar Yadav}
\affiliation{Aryabhatta Research Institute of Observational Sciences (ARIES), Manora Peak, Nainital-263001, India}
\affiliation{Department of Applied Optics and Photonics, University of Calcutta, Kolkata, 700106, India}
\email{sarvesh@aries.res.in}

\begin{abstract}

The \texttt{PyLMT} transient detection pipeline has been operational since November 2023, detecting transient and variable objects in the ILMT images in almost real time. Using the image subtraction technique, nearly 3700 CCD frames have been analyzed by the automated pipeline, generating $\sim$ 23,000 alerts for the detection of verified transient candidates and cataloged variable sources. Around 21,000 of the alerts correspond to known MPC asteroids, $\sim$ 2000 correspond to variable stars (including eclipsing binaries, RR Lyrae, Delta Scuti, T-Tauri, etc.), 509 correspond to variable AGNs (including QSOs, Seyfert galaxies, and blazars), 21 supernova candidates, and several other interesting candidates. We provide a concise overview of the detections and their significance, emphasizing the survey's potential contributions to a broad class of astrophysical and scientific cases. A transient detection dashboard called \texttt{DART} was developed using \texttt{Streamlit} to visualize and categorize candidates based on \texttt{PyLMT} and \texttt{SIMBAD} classifications. It includes cone-search functionality and displays key metadata, offering an intuitive interface that is publicly accessible. Our results demonstrate the viability of liquid mirror telescopes such as the ILMT for time-domain astronomy, emphasizing the important role that small-field survey facilities can play in systematic transient science programs.

\end{abstract}



\section{Introduction}
\label{sec:introduction}

Astrophysical transients and variable phenomena are some of the most important subjects of investigation in the era of modern time-domain astronomy. Such events are associated with a wide diversity in origins, ranging from Galactic sources, including a diverse population of variable stars, to extragalactic sources, including supernovae (SNe) and accretion-- or jet--driven variability like that in active galactic nuclei \citep[AGN;][]{2001sac..conf....3P}. SNe are broadly classified into thermonuclear and core-collapse events. Type Ia SNe or thermonuclear SNe \citep{2000ARA&A..38..191H, 2014ARA&A..52..107M}, arise from the explosion of white dwarfs through either single-- or double--degenerate channels, whereas all other SN subtypes originate from the core collapse of massive stars \citep{1997ARA&A..35..309F,2009ARA&A..47...63S}. On the galactic scales, stellar variability broadly manifests as eclipsing, rotational, pulsational, eruptive, cataclysmic, and secular variables \citep{2008JPhCS.118a2010E}. The optical transient sky encompasses phenomena evolving over timescales ranging from minutes to years, from rapidly evolving fast optical transients to slowly evolving SNe and long-term variability in AGN. Continued discovery and monitoring of these sources provide important constraints on stellar evolution, stellar explosions, accretion physics, and cosmology.

The advent of dedicated time-domain surveys has revolutionized the discovery and characterization of transient and variable sources by repeatedly imaging large regions of the sky. Major surveys, including the Palomar Transient Factory \citep[PTF;][]{2009PASP..121.1395L}, the Zwicky Transient Facility \citep[ZTF;][]{2019PASP..131a8002B}, the Asteroid Terrestrial-impact Last Alert System \citep[ATLAS;][]{2018PASP..130f4505T}, the Panoramic Survey Telescope and Rapid Response System \citep[Pan-STARRS;][]{2016arXiv161205560C}, the Gravitational-wave Optical Transient Observer \citep[GOTO;][]{10.1093/mnras/stac013} etc., have transformed time-domain astronomy by generating unprecedented volumes of transient alerts and variable-source detections. The Vera C. Rubin Observatory's Legacy Survey of Space and Time \citep[LSST;][]{2019ApJ...873..111I} is expected to further supersede the volume of transient alerts in the future.

The 4-m International Liquid Mirror Telescope\footnote{\url{https://www.aries.res.in/facilities/astronomical-telescopes/ilmt}}\textsuperscript{,}\footnote{\url{http://www.ilmt.ulg.ac.be/home/}} \citep[ILMT;][]{2025A&A...694A..80S} is the first optical survey telescope in India and the only operational astronomical liquid mirror telescope in the world. Operating in the time-delay integration (TDI) mode, the ILMT repeatedly surveys the same zenithal strip of the sky, making it well suited for the discovery of optical transients and long-term monitoring of variable sources. To enable efficient processing of the survey data, we developed the automated transient detection pipeline, \texttt{PyLMT} \citep{2025MNRAS.538..133P}, which performs image subtraction, candidate extraction, machine-learning (ML)-assisted classification, and cross-matching with archival catalogs to identify astrophysical transients and variable sources. The ML models are routinely updated to optimize pipeline performance \citep{10.1093/rasti/rzag044}.

In this work, we present a census of the transient and variable population detected with the ILMT between November 2023 and May 2025. Owing to the telescope's observing strategy, cadence, limiting magnitude, and field of view, the diversity of detectable sources is naturally constrained. Consequently, rather than providing an extensive review of all known classes of astrophysical transients and variable phenomena, this work focuses on the populations recovered by the ILMT survey and summarizes their observed characteristics.

The paper is organized as follows. Section~\ref{sec:Observations} is dedicated to details about data acquisition with the ILMT and archival data obtained from the ZTF. The diversity and nature of the detected transient and variable population are discussed in Section~\ref{sec:Detected Populations}. A user-friendly web-based framework developed to make the detected transient and variable alerts publicly available is discussed in Section~\ref{sec:DART}. Finally, the broader implications of this work for similar transient search campaigns are discussed, along with a summary, in Section~\ref{sec:Discussion}.

\section{Data acquisition} 
\label{sec:Observations}

\subsection{ILMT data acquisition}

ILMT is a 4-m aperture zenith-pointing telescope with f-ratio $\sim$ \textit{f}/2.36. The primary reflecting mirror is a rotating bowl filled with $\sim$50 liters of mercury. The balance between centrifugal force and gravity maintains the parabolic shape of the mercury surface. The bowl is a carbon fiber-epoxy skin over a closed foam core, mounted on an air-bearing and supported by a 3-point mount system. The mirror is surrounded by a metallic structure that houses the optical corrector and a 4K$\times$4K CCD detector. The telescope's effective FoV is $22'.4\times22'.4$ ($\sim$ 0.14 deg$^2$). A relatively large aperture enables the detection of faint sources of magnitudes up to $\sim$22 in the SDSS \textit{g}$'$ band. 

Due to the telescope's fixed pointing, the time-delay integration (TDI) technique is used for imaging. This technique compensates for the sidereal motion of the sky to deliver stationary images. The CCD remains fixed and is read out at the exact sidereal rate along the east-west direction. The continuous readout ensures propagation of accumulated charges at the same sidereal rate, thereby counterbalancing star trails and rendering point-like stellar images. This technique was first implemented with the Steward Observatory 1.8-m CCD/Transit instrument \citep[CTI;][]{10.1117/12.933448}. It has been observed that the TDI imaging suffers from a north-south elongation at non-zero latitudes due to conical trajectories of stellar images on the CCD plane \citep{1992MNRAS.258..543G}. The optical corrector affixed in front of the focal plane ensures that the conical trajectory traversed by the sources on the CCD plane is made rectilinear \citep{1998PASP..110.1081H,2024BSRSL..93..863N}. The parameters of the ILMT are given in Table~\ref{tab:ILMT_parameters}.

The ILMT observes the same local sidereal time (LST) fields on successive nights, with a 3-minute 56-second shift. The CCD integration time per source is 102.36 s, corresponding to the time taken for an object to traverse the $4096 \times 4096$ pixel detector operated in continuous readout mode. In TDI mode, ten such integrations occur continuously along the scan direction, resulting in a total exposure time of $\sim$17 minutes. However, since full integration time is reached only after the first 4096 rows are read, these initial rows are discarded. The resulting science image spans 1.25 $\mathrm{deg}^2$ with dimensions of $36864 \times 4096$ pixels. Images are acquired in a single filter per night. Depending on the season, up to 35 science frames can be acquired on a typical observation night with favorable weather conditions. This corresponds to up to $\sim$ 44 $\mathrm{deg}^2$, while the average sky coverage is $\sim$ 36 $\mathrm{deg}^2$. These images are processed using the \texttt{PyLMT} pipeline to search for astronomical transients, asteroids, and other variable sources. At the time of writing this paper, four observing cycles have concluded: October--November 2022, March--June 2023, November 2023--May 2024, and October 2024--May 2025.  

\begin{table}
\centering
\caption{Telescope and detector parameters of the ILMT \protect\citep{BKumar2022}.}
\begin{tabular}{cc}
\hline
Parameter & Value \\
\hline
Aperture size & 4.0-m diameter \\
f-ratio &  $\sim$2.36 \\ 
Field of view & $22'.4$$\times$$22'.4$ ($\sim$ 0.14 deg$^2$) \\
Accessible sky area &  $\sim$36 degree\textsuperscript{2} per night \\
Bowl rotation period & 8.02 sec \\
CCD Size & 4096$\times$4096 pixels \\
Pixel size & 0$''$.328 pixel\textsuperscript{-1} \\
Readout noise & 5.0 e$^{-}$ \\
Gain & 4.0 e$^{-}$/ADU \\
Integration time & 102.36 sec \\
Filters & SDSS \textit{g}$'$, \textit{r}$'$, \textit{i}$'$ \\
\hline
\end{tabular}
\label{tab:ILMT_parameters}
\end{table}

\subsection{Archival ZTF data}
\label{sec:ZTF}

ILMT observations provide deep, nightly imaging of a fixed zenith strip with stable observing conditions and well-defined photometric characteristics. However, owing to the fixed pointing geometry and the limited temporal baseline of individual observing cycles, ILMT lightcurves alone can be insufficient for robust period determination or long-term variability characterization. To mitigate this limitation, ZTF archival photometry is incorporated whenever available to extend the temporal coverage of ILMT detections.

For periodic and quasi-periodic variables such as eclipsing binaries and RR Lyrae stars, ZTF data are used to determine or refine variability periods and to construct phase-folded lightcurves. In these cases, the ILMT detections establish the presence of variability and provide deep multi-band measurements, while the ZTF data supply dense temporal sampling over multi-year baselines. This combined approach enables reliable classification and, for some newly identified variable sources, detailed lightcurve modelling.

For transient sources, including SN candidates and eruptive variables, ZTF lightcurves are used to validate ILMT detections and place them in a broader temporal context. Archival ZTF data can enable the identification of past outbursts, variability, confirmation of transient behaviour, and discrimination between explosive events and long-term variable sources such as AGN. In several cases, ZTF photometry samples phases of the transient evolution that are not covered by the ILMT observations.

\section{Transient and Variable Source Population identified with ILMT} \label{sec:Detected Populations}

The objects detected with the ILMT represent a diverse population of astrophysical phenomena (Table~\ref{tab:summary_classification}). The majority of detections correspond to asteroids and small Solar System bodies, reflecting the survey’s continuous coverage and sensitivity to moving objects. Variable stars form the next largest group, dominated by eclipsing binaries and RR Lyrae, followed by a smaller but significant population of AGN (QSOs, Seyferts, and related sources). In addition, rarer but astrophysically valuable events such as cataclysmic variables (CVs)/novae, X-ray sources (which can include sources with detected X-ray counterparts but no accurate determination of physical nature, and could belong to astrophysical types like AGNs, and CVs, etc.), and SN candidates are also identified. This diversity highlights ILMT’s ability to probe astrophysical variability across a wide range of scales, from Solar System dynamics to stellar evolution and extragalactic activity. A subset of these categories is discussed in detail in the following subsections.

\begin{table*}[ht]
\centering
\caption{Summary of \texttt{PyLMT} detections in different categories of astrophysical transients and variable sources. The `Object' column represents the total number of different objects detected in that category. `All Detections' column represents the total number of detections for all objects in that category.}
\label{tab:summary_classification}
\begin{tabular*}{\textwidth}{l@{\extracolsep{\fill}}cc}
\hline
\textbf{Category} & \textbf{Objects} & \textbf{All Detections} \\
\hline
\multicolumn{3}{l}{\textit{AGNs}} \\
\hspace{1em}QSO & 105 & 403 \\
\hspace{1em}Sy1 & 23 & 101 \\
\hspace{1em}Blazar & 3 & 5 \\
\textbf{Subtotal AGNs} & \textbf{131} & \textbf{509} \\
\hline
\multicolumn{3}{l}{\textit{Variable Stars}} \\
\hspace{1em}EB* & 339 & 1322 \\
\hspace{1em}RR* & 90 & 409 \\
\hspace{1em}CV/Novae & 7 & 46 \\
\hspace{1em}Other Variables & 83 & 207 \\
\textbf{Subtotal Variables} & \textbf{519} & \textbf{1984} \\
\hline
\textbf{Supernova candidates} & \textbf{21} & \textbf{186} \\
\textbf{Asteroids/small bodies} & \textbf{9034} & \textbf{20732} \\
\hline
\textbf{TOTAL} & \textbf{9705} & \textbf{23411} \\
\hline
\end{tabular*}
\end{table*}

\subsection{Transients/SNe}

SNe are a class of extragalactic astronomical transient events that can occur either due to the core collapse of massive stars with masses greater than approximately 8 $M_{\odot}$ \citep{2009ARA&A..47...63S,2015PASA...32...16S}, or due to the thermonuclear disruption of a white dwarf triggered by binary accretion \citep{2007Sci...315..825M,2013IAUS..290..117H}. Although significant progress has been made in understanding SNe, uncertainty remains about their progenitor properties, environments, and explosion mechanisms. The advent of sky surveys has enabled the detection of many of these events, helping to bridge gaps in our understanding. An aperture size of 4 m, a daily cadence, and zenith sky observations with minimal atmospheric extinction enable the ILMT to detect and follow up bright and faint SN candidates.

The \texttt{PyLMT} pipeline processes images acquired during observations in near real time. The team manually vets the candidate list the next day, and those classified as \textit{extended-host} are separated as potential SN candidates. This classification is important because it further increases the likelihood of an event having a clear host-galaxy association, as is the case with many SNe. Although the pipeline separates out the asteroid detection, the candidate is manually re-checked with the Minor Planet Center (MPC) checker tool\footnote{\url{https://www.minorplanetcenter.net/cgi-bin/checkmp.cgi}} to confirm the absence of a known asteroid. The candidate is cross-matched with public databases like \texttt{SIMBAD}, \texttt{VizieR} \citep{2000A&AS..143...23O}, ZTF alerts \citep{2024TNSTR.405....1F}, and AAVSO \citep{2006SASS...25...47W} to reject any known sources of variability (\textit{viz.} variable stars, AGNs, and novae, etc). Finally, it is also ensured that the candidate is present in multiple ILMT frames to further reduce the possibility of an artefact or a non-cataloged asteroid. The SN candidates are then cross-matched with the transient name server (TNS)\footnote{\url{https://www.wis-tns.org/}} database. The detected candidate is reported to the TNS as a detection with the ILMT magnitudes (Dukiya et al. 2026; under preparation). If the candidate is not already present in the database, it is reported as a discovery. Table~\ref{tab:ilmt_sne} lists all the ILMT-detected and/or discovered transient candidates reported to the TNS.

\begin{figure*}[ht]
    \centering
\fbox{%
    \includegraphics[width=0.48\textwidth]{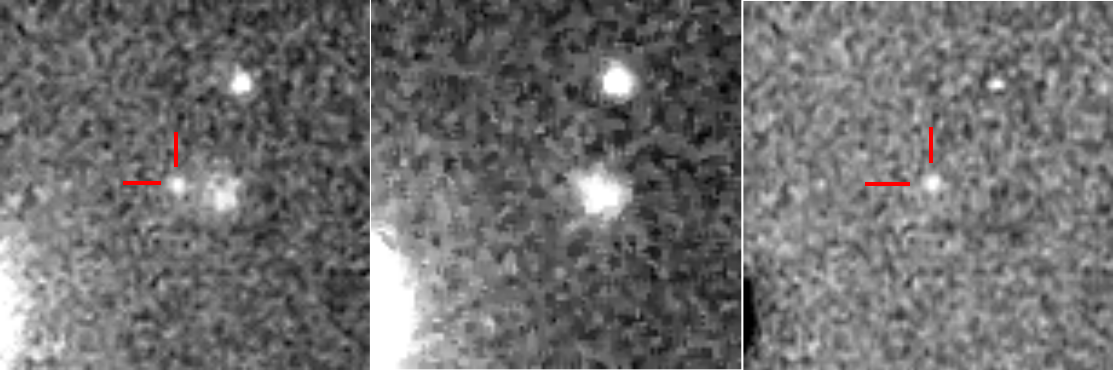}%
}
\fbox{%
    \includegraphics[width=0.48\textwidth]{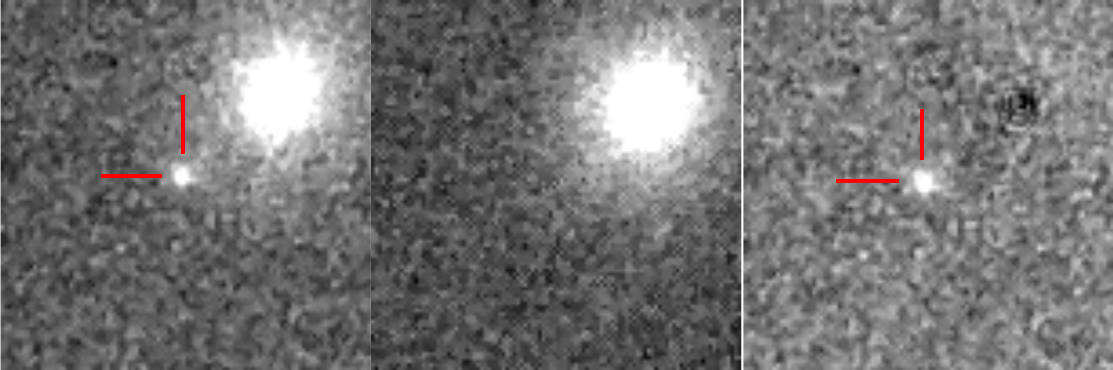}%
}

\vspace{1pt}

\fbox{%
    \includegraphics[width=0.48\textwidth]{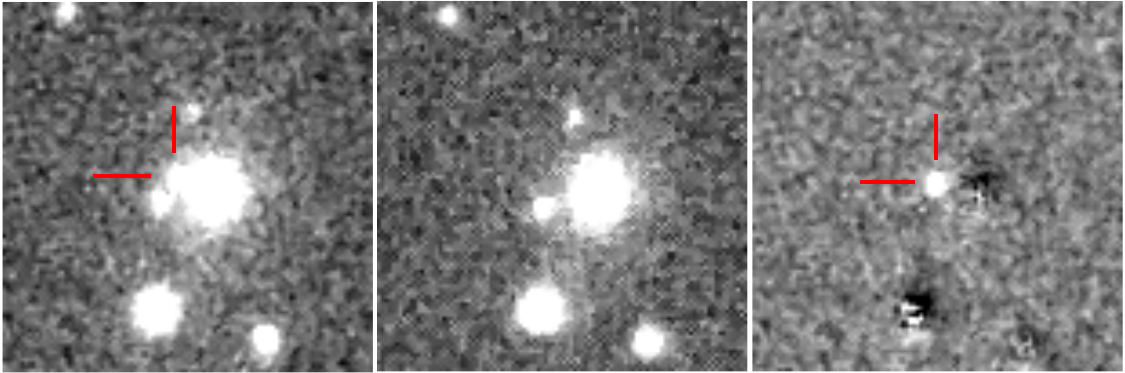}%
}
\fbox{%
    \includegraphics[width=0.48\textwidth]{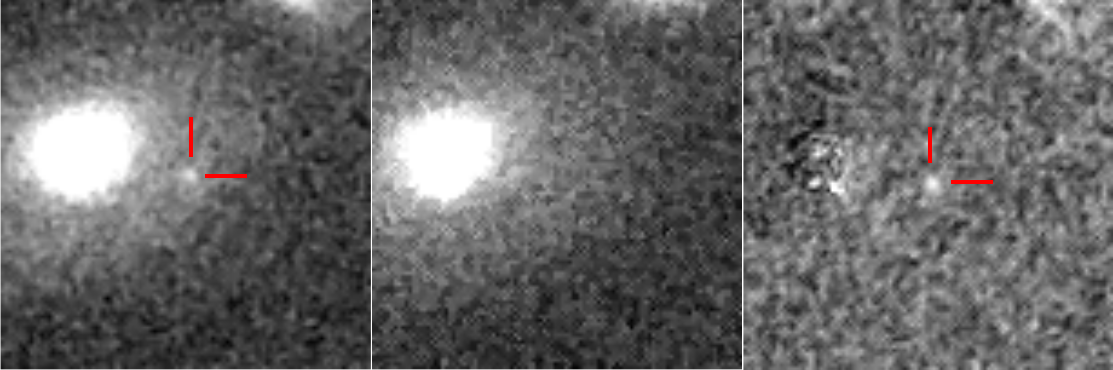}%
}

\vspace{1pt}

\fbox{%
    \includegraphics[width=0.48\textwidth]{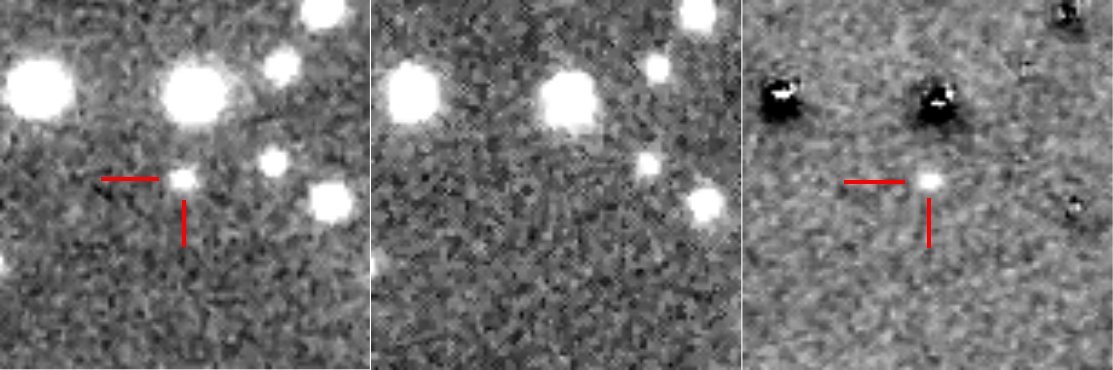}%
}
\fbox{%
    \includegraphics[width=0.48\textwidth]{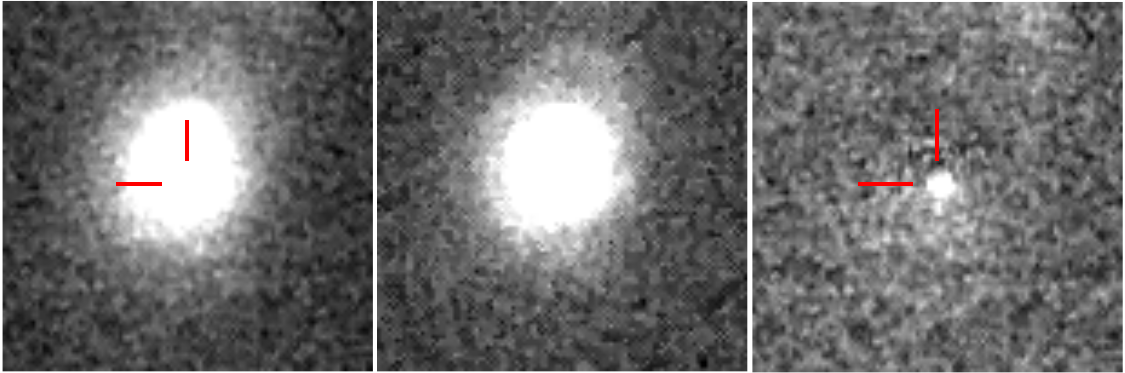}%
}
    \caption{From left to right in each panel: science, reference, and difference image cutouts of ILMT-discovered transient candidate. The left column shows (top to bottom) AT~2023yjc, 2024zsm, and 2024aiha. The right column shows (top to bottom) AT~2024fxn, 2024aifv, and 2024agkc.}
    \label{fig:SN_cutouts}
\end{figure*}

\begin{figure*}
    \centering
    \begin{subfigure}[b]{0.47\textwidth}
        \centering
        \includegraphics[width=1.1\textwidth]{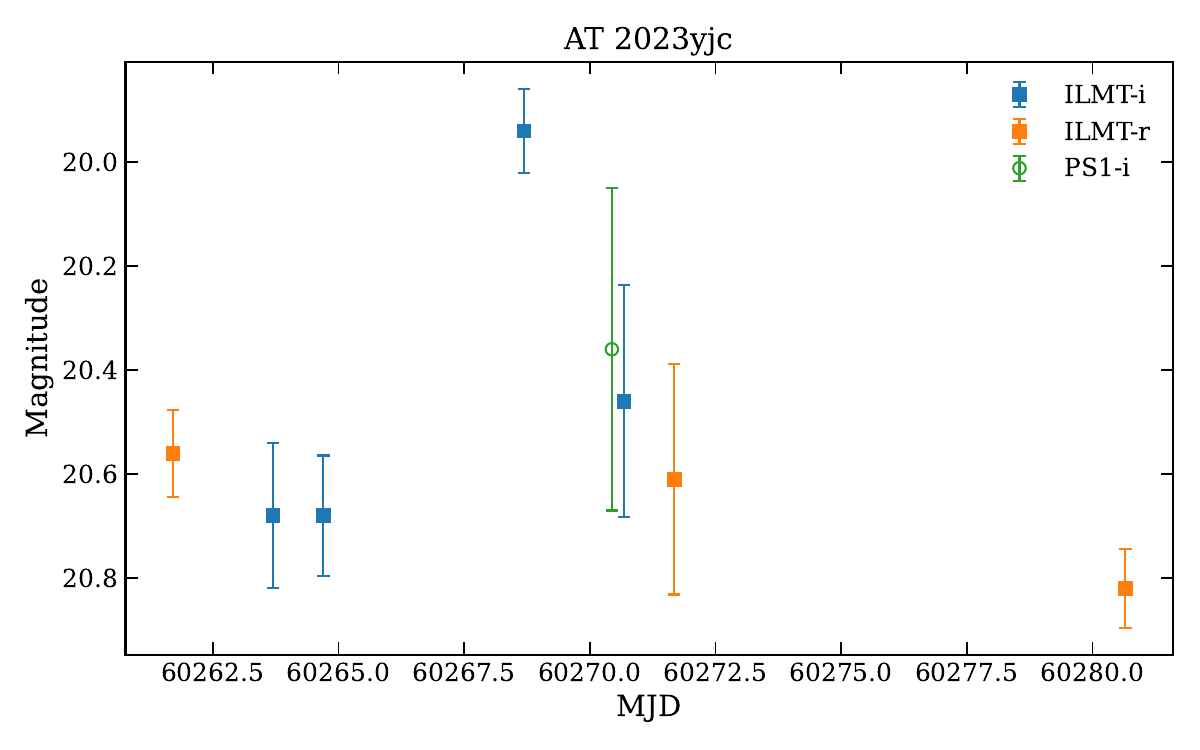}
        \label{fig:AT2023yjc}
    \end{subfigure}
    \hfill
    \begin{subfigure}[b]{0.47\textwidth}
        \centering
        \includegraphics[width=1.1\textwidth]{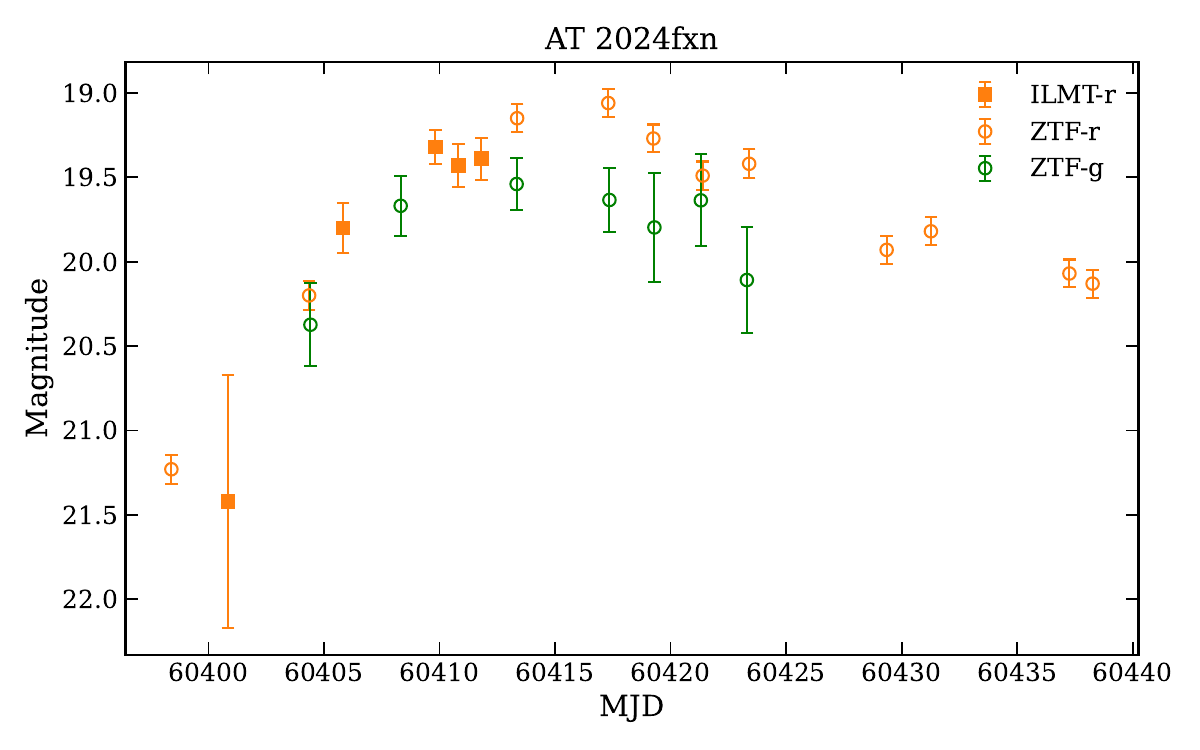}
        \label{fig:AT2024fxn}
    \end{subfigure}%
    \hfill
    \begin{subfigure}[b]{0.47\textwidth}
        \centering
        \includegraphics[width=1.1\textwidth]{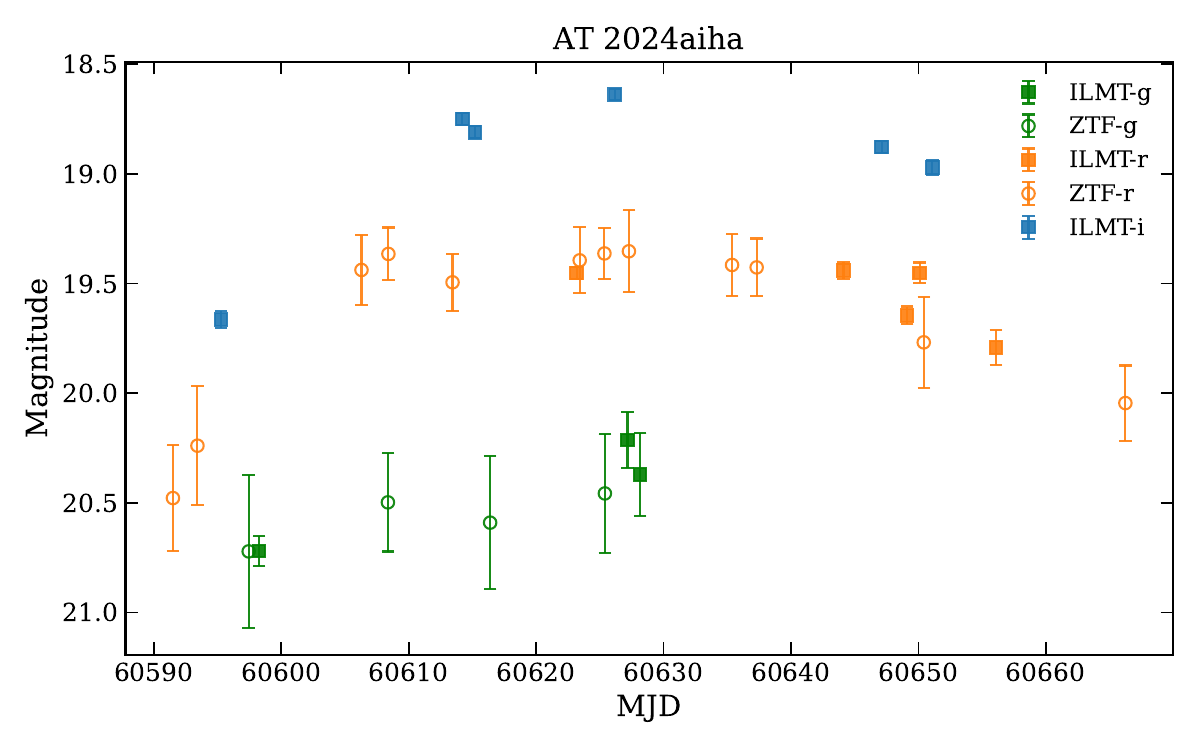}
        \label{fig:AT2024aifv}
    \end{subfigure}
    \hfill
    \begin{subfigure}[b]{0.47\textwidth}
        \centering
        \includegraphics[width=1.1\textwidth]{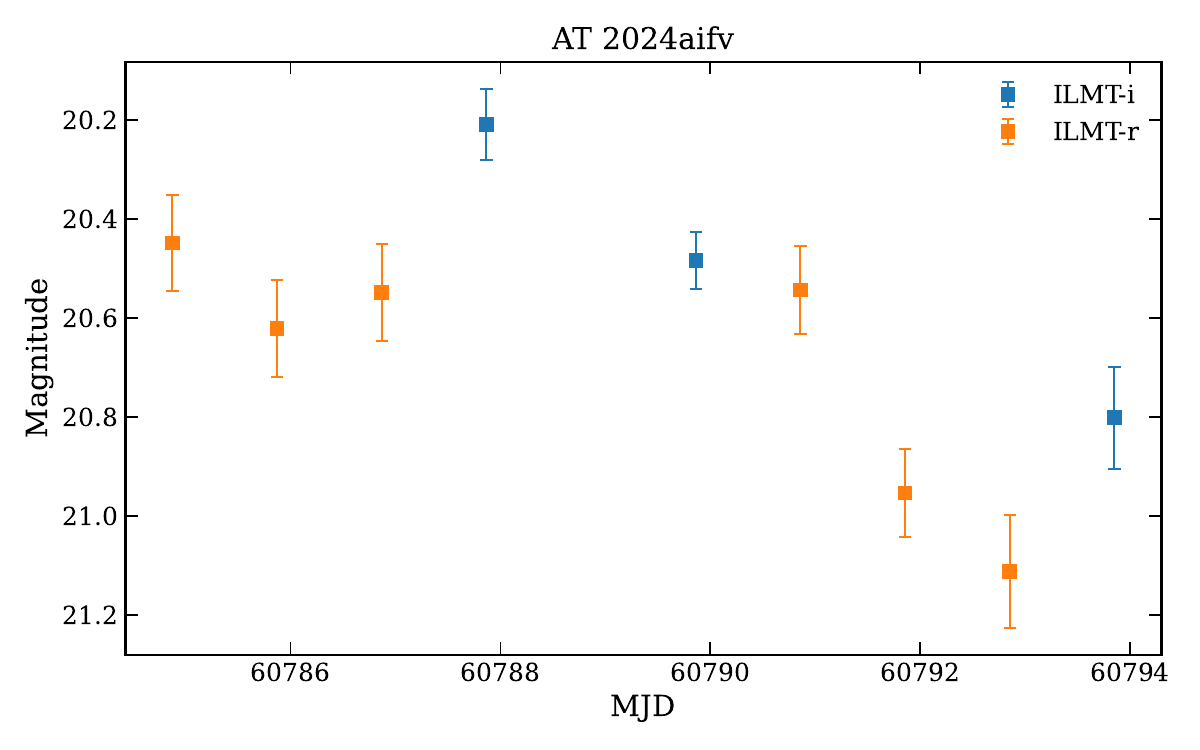}
        \label{fig:AT2024aiha}
    \end{subfigure}%
    \caption{Clockwise: Multi-band lightcurves of three SN candidates -- AT~2023yjc, 2024fxn, 2024aifv, and one ambiguous transient source 2024aiha, discovered with the ILMT.}
    \label{fig:SN_discovered}
\end{figure*}

Operational since November 2023, the pipeline has identified 21 SN-like transient candidates. Five of the detected SN candidates, viz. AT~2023yjc \citep{2023TNSTR3062....1P}, 2024fxn \citep{2024TNSTR.964....1P}, 2024zsm \citep{2024TNSTR4208....1P}, 2024agkc \citep{2025TNSTR..20....1P}, and 2024aifv \citep{2025TNSTR1468....1P} were reported to the TNS as discoveries. In addition to these 5 candidates, there was a detection of one ambiguous transient candidate AT~2024aiha \citep{2025TNSTR3845....1P}, whose classification could not be securely established. These candidates were relatively faint, with SDSS \textit{r}$'$ and \textit{i}$'$ magnitudes ranging from 19 to 21 mag, which prevented spectroscopic confirmation and classification. The remaining 14 candidates, first discovered by other transient surveys, were also independently detected with the ILMT and some of them were reported to the TNS. Figure~\ref{fig:SN_cutouts} shows the discovery science, reference, and difference images of all five ILMT-discovered SN candidates and the one ambiguous candidate. With the exception of AT~2024aiha, the other 5 candidates show clear host galaxies, further strengthening their SN origin. The lightcurves are also presented for three out of five SN candidates with good photometric coverage and the one ambiguous case in Figure~\ref{fig:SN_discovered}. The ILMT photometry is complemented by publicly available measurements from the ZTF survey. 
The five discovered candidates and the one ambiguous event are discussed in more detail below. Also, out of the remaining 14 detected but not discovered candidates, a spectroscopic follow-up study was performed for SN~2024cjb, which is discussed in more detail below.

\subsubsection*{AT~2024{\lowercase{fxn}}}

AT~2024fxn ($\mathrm{RA} = 14^{\mathrm{h}}13^{\mathrm{m}}49.54^{\mathrm{s}},\; \mathrm{Dec} = +29^\circ 23' 32.4''$) was discovered on 5 April 2024 \citep{2024TNSTR.964....1P, 2026arXiv260515693F}. It was detected as a transient source near the outskirts of a galaxy that appears to have an early-type (elliptical-like) morphology in the available imaging. The r-band light curve shows a clear rise in brightness from r-mag $\sim$21.2 mag at MJD~60398 to a peak magnitude of r-mag $\sim$19.1 mag by MJD~60416, corresponding to a brightening of nearly 2~mag over almost 17~days. Following maximum light, the source exhibits a steady decline, fading from r-mag $\sim$19.1 mag at MJD~60416 to r-mag $\sim$19.8 mag by MJD~60431, corresponding to a decline of nearly 0.7$\pm$0.1~mag over almost 15~days. The overall lightcurve evolution, characterized by a smooth rise and subsequent decline on timescales of weeks, is consistent with an SN candidate. Additionally, the early-type host morphology (Figure \ref{fig:SN_cutouts}) supports a Type Ia origin for this SN candidate. However, the absence of spectroscopic observations prevents a secure classification of the event.

\subsubsection*{AT~2024{\lowercase{aiha}}}

AT~2024aiha ($\mathrm{RA} = 05^{\mathrm{h}}39^{\mathrm{m}}53.40^{\mathrm{s}},\; \mathrm{Dec} = +29^\circ 29' 04.0''$), first detected on 10 October 2024 \citep{2025TNSTR3845....1P}, is a peculiar transient. No host galaxy was identified in the reference images, so the pipeline flagged the source as a hostless candidate. However, follow-up analysis using co-added ILMT images revealed a faint underlying host. The transient lies very close to the Galactic plane ($b = -0.77^\circ$), implying substantial Galactic extinction, which may also hinder detection of its faint host. The gri-band lightcurves (Figure~\ref{fig:SN_discovered}; not corrected for extinction) exhibit a smooth rise and subsequent decline, consistent with a transient source, and hence motivated its reporting to the TNS. The source brightened from $r \sim 20.5$ mag to $r \sim 19.3$ mag over $\sim 33$ days. After peak, the decline is gradual, fading by only $\sim 0.2 \pm 0.05$ mag over $\sim 15$ days. 

The broad morphology of the light curve resembles that of a Type Ic supernova. However, after correcting for Galactic extinction, the observed $g-r$ color is substantially bluer than expected for such an event. Adopting $E(B-V)=1.482 \pm 0.06$ mag \citep{2011ApJ...737..103S} and $R_V=3.1$, the corresponding extinctions are $A_g=4.896$ mag and $A_r=3.387$ mag. The extinction-corrected $g-r$ color evolves from approximately $-0.8$ at $\sim6$ days to $-0.4$ at $\sim33$ days after the first detection. This color evolution is not typically observed in the major supernova classes \citep{2002PASP..114..833P}, and therefore does not support a supernova interpretation. It should be mentioned that any color-based examination should be treated with caution due to systematic uncertainties associated with high extinction values in such regions. An alternate possibility is that the source is a nova-like galactic transient. However, no cataloged galactic nova source (or other transient or variable source) could be crossmatched in the \texttt{VizieR} or International Variable Star Index \citep[VSX;][]{2006SASS...25...47W}\footnote{\url{https://vsx.aavso.org/index.php?view=search.top}} catalog at that position. Considering the above facts and the absence of spectroscopic data, the source could not be classified into any well-defined class, and its nature remains ambiguous.   

\subsubsection*{AT~2023{\lowercase{yjc}}, 2024{\lowercase{zsm}}, 2024{\lowercase{agkc}}, \lowercase{and} 2024{\lowercase{aifv}}}

\textbf{AT~2023yjc} was the first supernova candidate discovered with the ILMT. It was detected on 13 November 2023 at $\mathrm{RA} = 01^{\mathrm{h}}50^{\mathrm{m}}02.90^{\mathrm{s}},\; \mathrm{Dec} = +29^\circ 08' 53.0''$. The detection magnitude was estimated to be r-mag $\sim$20.6 mag. The lightcurve of the transient source is presented in Figure~\ref{fig:SN_discovered}.

\textbf{AT~2024aifv} was discovered with the ILMT on 19 April 2025 at $\mathrm{RA} = 16^{\mathrm{h}}12^{\mathrm{m}}11.10^{\mathrm{s}},\; \mathrm{Dec} = +29^\circ 34' 17.0''$. The detection magnitude of the source was determined to be r-mag $\sim$20.3 mag. The lightcurve of the transient source is shown in Figure~\ref{fig:SN_discovered}.

\textbf{AT~2024zsm} was discovered with the ILMT on 28 October 2024 at $\mathrm{RA} = 02^{\mathrm{h}}52^{\mathrm{m}}40.0^{\mathrm{s}},\; \mathrm{Dec} = +29^\circ 10' 09.0''$. The detection magnitude of the source was r-mag $\sim$20.2 mag. Insufficient data points prevented the construction of a detailed lightcurve for this source.

\textbf{AT~2024agkc} was discovered with the ILMT on 30 December 2024 at $\mathrm{RA} = 07^{\mathrm{h}}49^{\mathrm{m}}08.10^{\mathrm{s}},\; \mathrm{Dec} = +29^\circ 35' 07.0''$. The detection magnitude of the source was r-mag $\sim$19.5 mag. Again, insufficient data points prevented the construction of a detailed lightcurve for this source.

Detailed lightcurve characterization could not be performed for the candidates AT~2024yjc, 2024aifv, 2024zsm, and 2024agkc due to insufficient data. This is largely due to limited observations caused by weather constraints or the source becoming fainter than the telescope's detection limit. However, the presence of clear host galaxies in the detection images (see Figure \ref{fig:SN_cutouts}) of these sources, along with their multi-epoch detections (see Table \ref{tab:ilmt_sne}), supports their supernova origin. 

\begin{figure}[t]
    \centering

    \hspace*{-0.4cm}
    \fbox{%
        \includegraphics[width=0.48\textwidth]{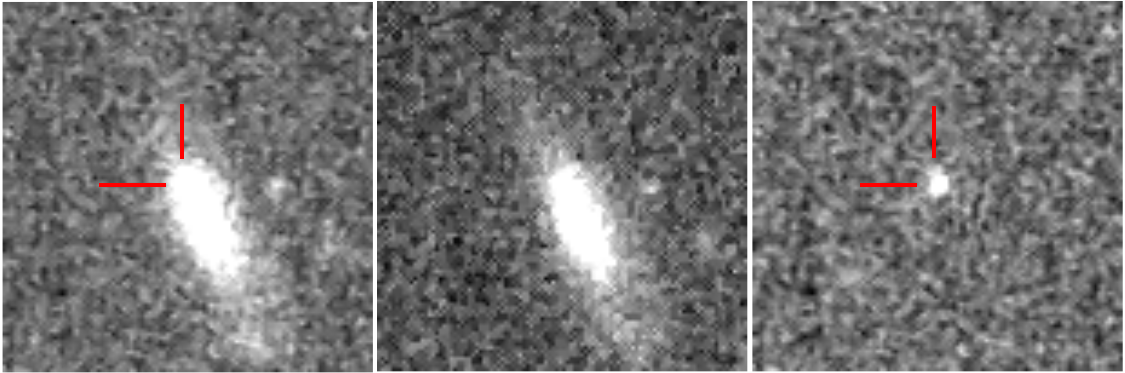}%
    }

    \par\vspace{0.1cm}
    (a)

    \vspace{0.3cm}

    \hspace*{-0.8cm}\includegraphics[width=\columnwidth]{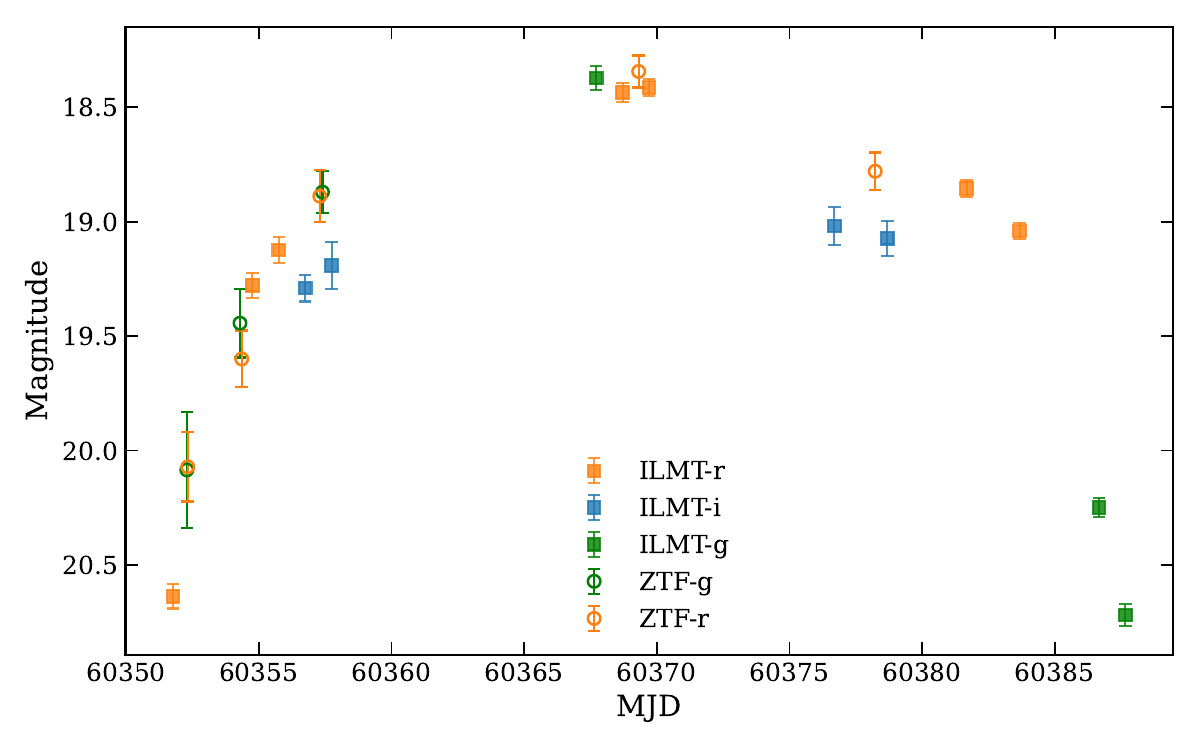}

    \par\vspace{0.1cm}
    (b)

    \vspace{0.3cm}

    \hspace*{-0.8cm}\includegraphics[width=\columnwidth]{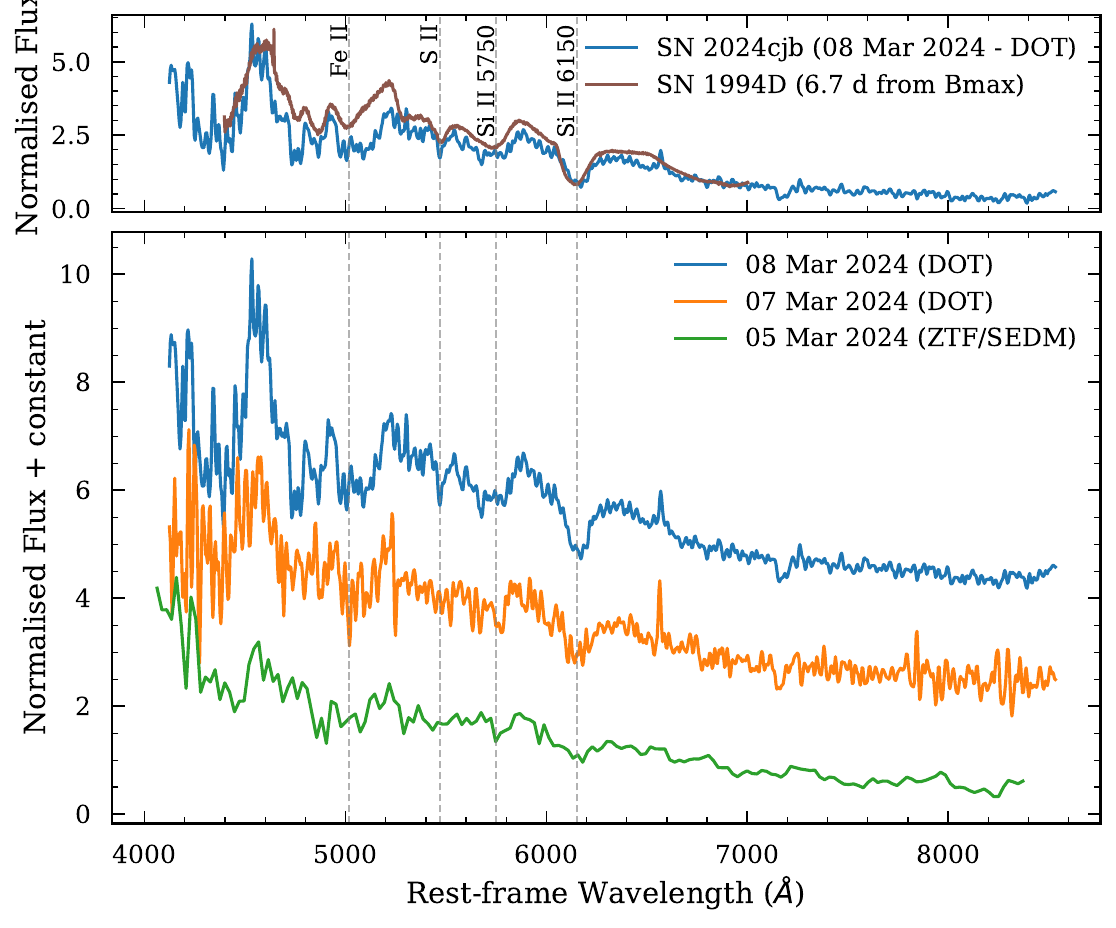}

    \par\vspace{0.1cm}
    (c)

    \caption{Overview of SN~2024cjb. (a) Science, reference, and difference image cutouts of the Type Ia SN~2024cjb detected with the ILMT on 14 February 2024. (b) Multi-band light curve of SN~2024cjb. (c) Spectra of SN~2024cjb obtained with DOT (ADFOSC) and ZTF (SEDM) at different epochs. A best-matching \texttt{GELATO} supernova template spectrum is also shown for comparison.}
    \label{fig:SN2024cjb_overview}
\end{figure}

\subsubsection*{SN~2024{\lowercase{cjb}}}

SN~2024cjb ($\mathrm{RA} = 09^{\mathrm{h}}11^{\mathrm{m}}27.47^{\mathrm{s}},\; \mathrm{Dec} = +29^\circ 29' 36.5''$) was detected with the ILMT on 14 February 2024 (discovery images shown in Figure \ref{fig:SN2024cjb_overview} (subplot a)), 2 days after its original discovery with the ZTF \citep{2024TNSTR.405....1F}. It was classified as a Type Ia SN \citep{2024TNSCR.611....1J}. From MJD~60352 to 60364, it brightened by almost 2.4 mag (lightcurve shown in Figure \ref{fig:SN2024cjb_overview} (subplot b)), attaining a peak magnitude of around r-mag $\sim$18.3 mag. In the next 15 days, from MJD~60364 to 60379, the brightness declined by nearly $0.53\pm0.06$ mag. Low-resolution spectra were acquired with the ARIES Devasthal Faint Object Spectrograph \& Camera \citep[ADFOSC;][]{article_adfosc} instrument mounted on the 3.6m Devasthal Optical Telescope (DOT) on 7 and 8 March 2024, 25 and 26 days after the original discovery by the ZTF. The spectroscopic data were also complemented with spectra from the Palomar 60-inch telescope, made available on WiSeREP\footnote{\url{https://www.wiserep.org/}}. The spectra show a clear silicon absorption feature at 6150 \AA, consistent with its Type Ia nature. The lightcurve of SN~2024cjb is shown in the center panel of Figure~\ref{fig:SN2024cjb_overview}. To determine the spectral classification and evolutionary phase of SN~2024cjb, we employed the GEneric cLAssification TOol \citep[\texttt{GELATO};][]{2008A&A...488..383H} software, which compares an observed spectrum with a library of supernova spectra spanning different types and epochs to identify the best-matching template. The spectra of SN~2024cjb (bottom panel of subplot c), together with the best-fit comparison spectrum of the Type Ia SN~1994D returned by \texttt{GELATO} (top panel of subplot c), are shown in subplot c of Figure~\ref{fig:SN2024cjb_overview}.

\begin{figure*}
    \centering
    \begin{subfigure}[b]{0.48\textwidth}
        \centering
        \includegraphics[width=1\textwidth]{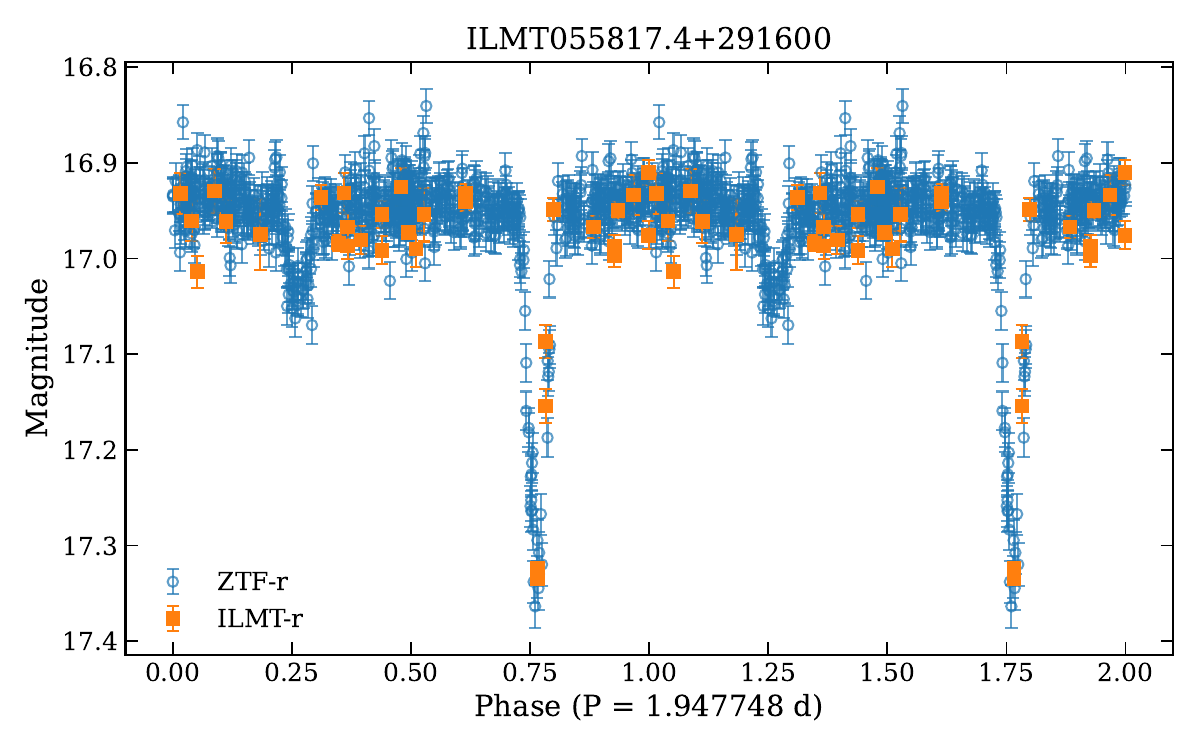}
        \label{fig:ILMT055817.4+291559}
    \end{subfigure}
    \hfill
    \begin{subfigure}[b]{0.48\textwidth}
        \centering
        \includegraphics[width=1\textwidth]{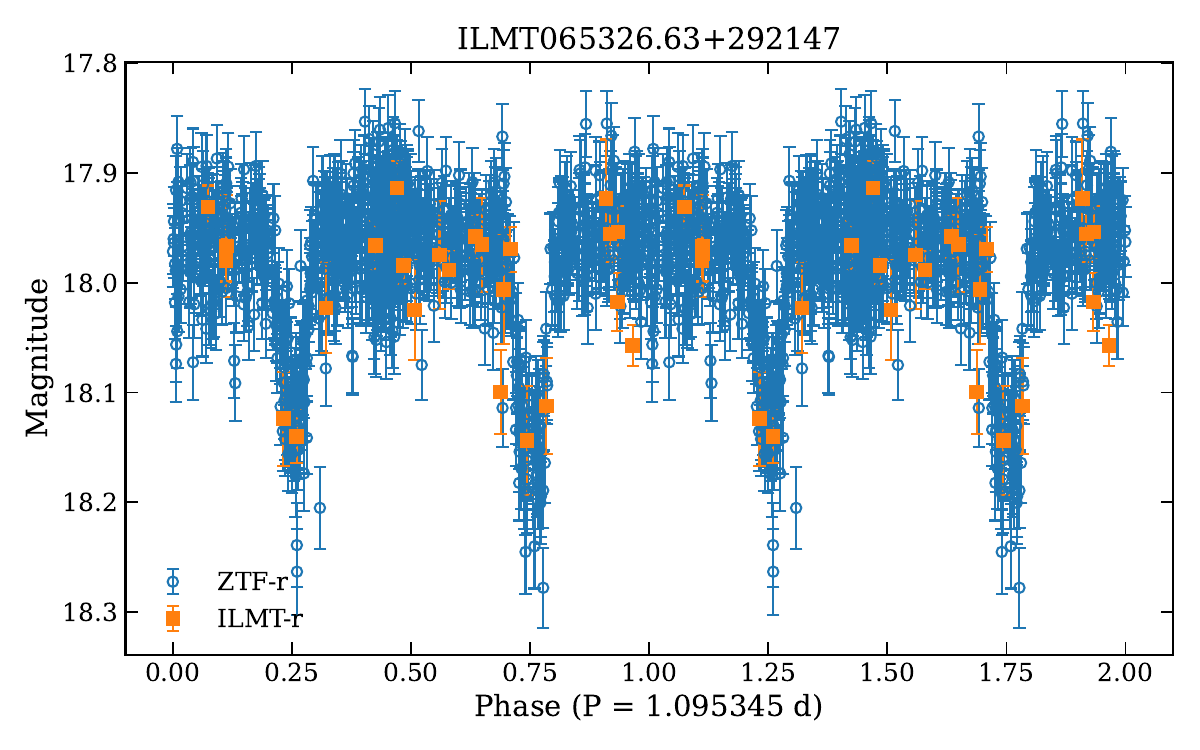}
        \label{fig:ILMT065326.63_292147}
    \end{subfigure}%
    \hfill
    \begin{subfigure}[b]{0.48\textwidth}
        \centering
        \includegraphics[width=1\textwidth]{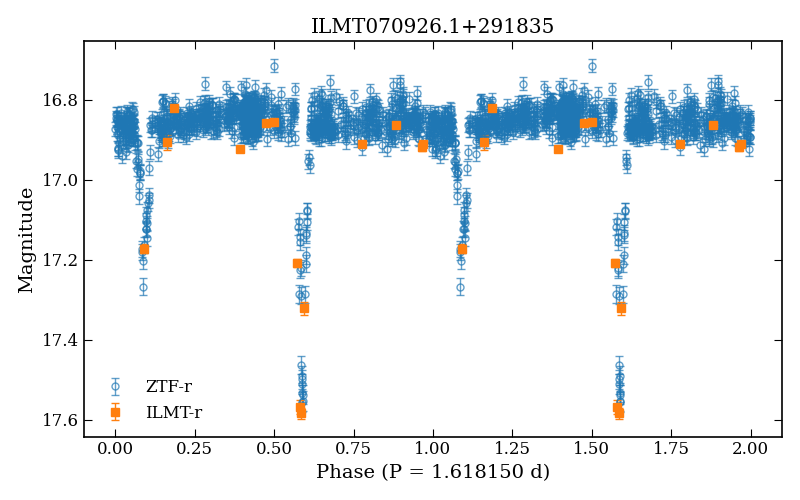}
        \label{fig:ILMT070926.1+291835}
    \end{subfigure}%
    \hfill
    \begin{subfigure}[b]{0.48\textwidth}
        \centering
        \includegraphics[width=1\textwidth]{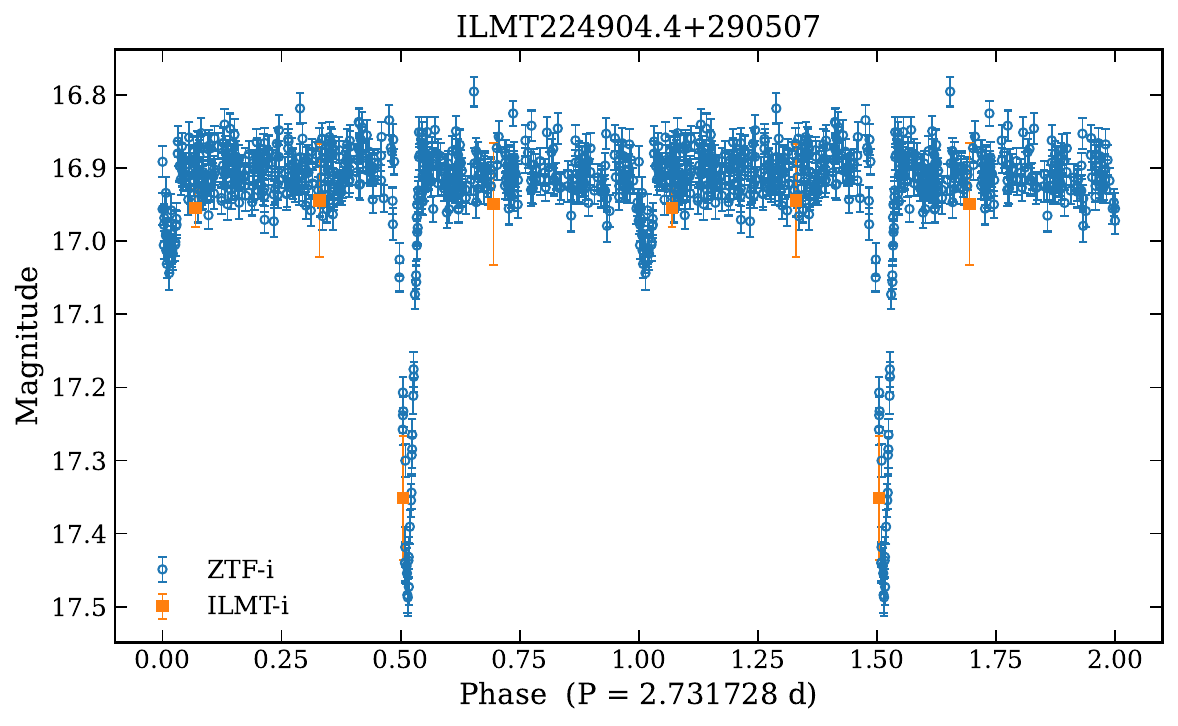}
        \label{fig:ILMT224904.4+290507}
    \end{subfigure}%
    \hfill
    \begin{subfigure}[b]{0.48\textwidth}
        \centering
        \includegraphics[width=1\textwidth]{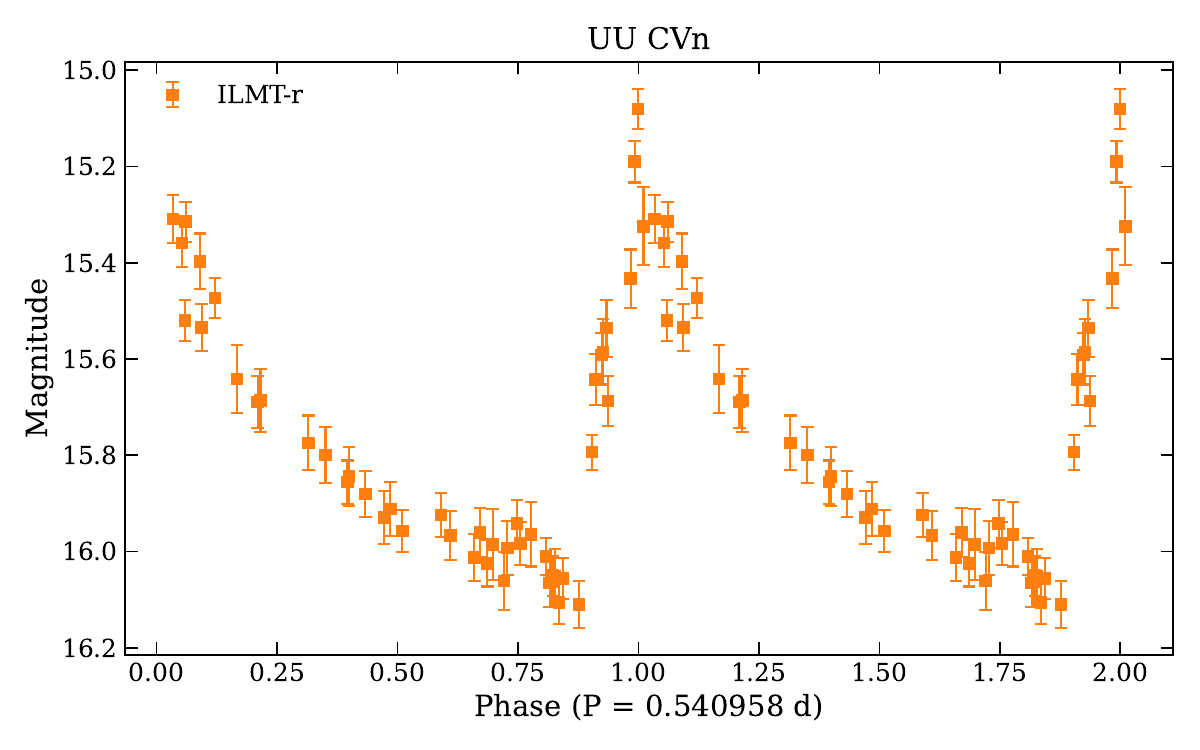}
        \label{fig:UU_CVn_lc_r}
    \end{subfigure}%
    \hfill
    \begin{subfigure}[b]{0.48\textwidth}
        \centering
        \includegraphics[width=1\textwidth]{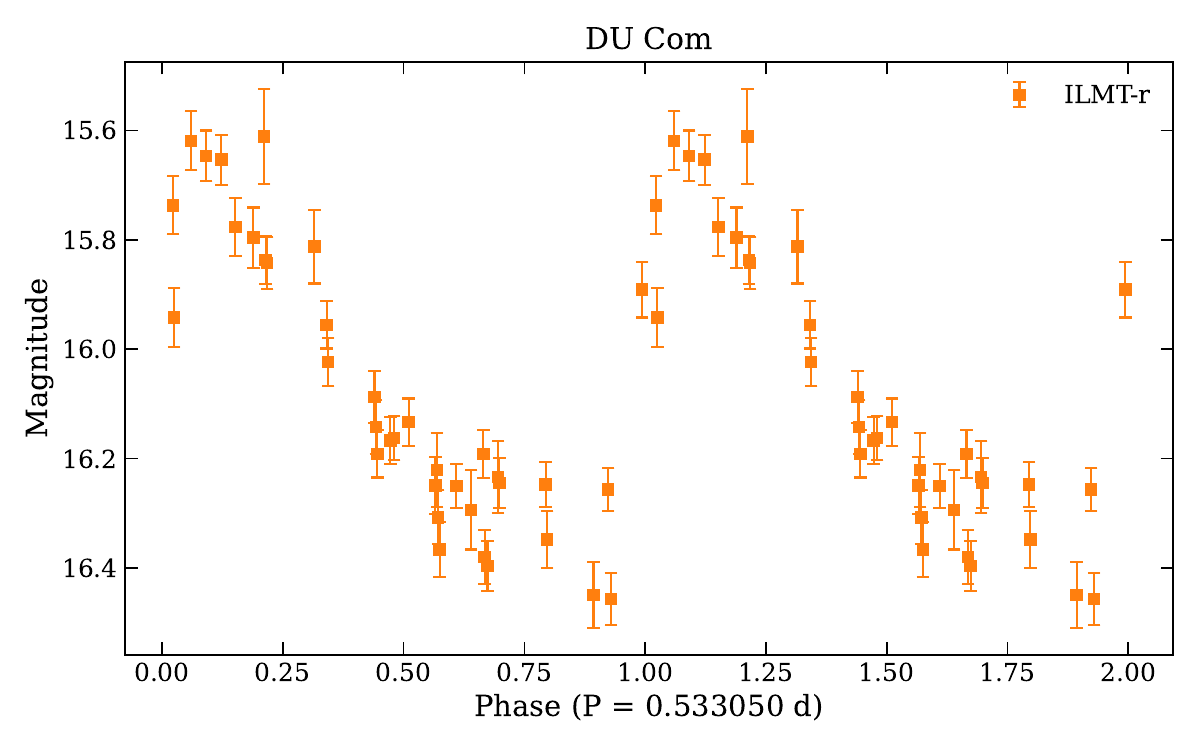}
        \label{fig:DU_Com_lc_r}
    \end{subfigure}%
    \caption{First two rows: Phase-folded lightcurves of four eclipsing binaries detected with the ILMT that are not present in the VSX catalog. Bottom row: Phase-folded lightcurves of two cataloged RR-Lyrae stars, UU CVn and DU Com, present in the ILMT FoV.} 
    \label{fig:new_var}
\end{figure*}

\subsection{Variable Stars \& Novae}

During the observation period, the zenith sky over Devasthal intersects the Galactic plane on two seasonal occasions. These passages provide repeated opportunities to monitor a rich field of galactic sources. In particular, this facilitates dedicated surveys of variable stars, novae, and other transient phenomena within the Milky Way, which can potentially contribute to studies of stellar variability and the understanding of nova outbursts. The faint limiting magnitude of ILMT enables the detection of variability in relatively faint variable star candidates. The \texttt{PyLMT} pipeline has detected more than 500 cataloged variable stars, including RR Lyrae, Delta Scuti, eclipsing binaries, and long-period variables. It has also detected several nova outbursts associated with CVs. Additionally, a list of variable star candidates absent in the VSX catalog maintained by the AAVSO was compiled. The sample includes candidates with at least two detections in ILMT data and at least ten detections in ZTF. The choice of this selection criterion is motivated by the fact that the ZTF survey has been running for a longer duration than the ILMT, hence a larger number of detections with the ZTF is naturally expected. To minimize contamination from AGN, these candidates were cross-matched with the \texttt{VizieR} and \texttt{SIMBAD} databases to ensure that they were not identified as quasars. Lightcurves combining ILMT and ZTF photometric data were then constructed, enabling further classification of the candidates into variability subtypes. It should be noted, however, that some objects from this list, although absent from VSX, had previously been reported as variable star candidates in \texttt{VizieR} through existing sky surveys, although they remained unclassified. The list of new variable star candidates, to the best of our knowledge at the time of submission of this draft, is presented in Table~\ref{tab:variable_sources}.

A detailed discussion of two major classes of variable stars (Eclipsing binaries and RR Lyrae) from this list is presented in Sections~\ref{sec:EBs} and \ref{sec:RR_Lyrae}. An elaborate discussion regarding two newly identified BY Draconis candidates is presented in Section \ref{sec:BY_Draconis}. A separate discussion on CV-associated variability is presented in Section~\ref{sec:CV_outbursts}.

\subsubsection{E\lowercase{clipsing Binaries}}
\label{sec:EBs}

In binary systems whose orbital planes are oriented close to the observer’s line--of--sight, mutual eclipses may occur when one component transits in front of the other, partially or completely occulting the flux from the eclipsed star. Such eclipsing binaries are identified by periodic, repeatable variations in the observed flux. Analysis of the resulting photometric lightcurves not only confirms the binarity of the system but also enables the determination of key stellar and orbital parameters. The detectability of an eclipsing binary (EB) using image differencing depends critically on the relative orbital phases at which the science and reference images are obtained.
Till the end of the fourth observing cycle (cycle of October 2024 -- May 2025), 9 new eclipsing binary candidates have been identified, which were not previously classified in VSX or \texttt{VizieR}. The lightcurves of four out of 9 newly identified candidates are shown in Figure~\ref{fig:new_var}. 

\subsubsection{RR Lyrae}
\label{sec:RR_Lyrae}

RR Lyrae stars are radially pulsating variables that occupy the horizontal-branch region of the classical instability strip in the Hertzsprung-Russell (H-R) diagram. These sources are the second-most frequently detected class of variable stars in this survey, after eclipsing binaries. The ILMT phase-folded lightcurves of two of the RR-Lyrae stars in the ILMT field are shown in Figure \ref{fig:new_var}. 

\begin{figure}
    \centering
    \includegraphics[width=0.48\textwidth]{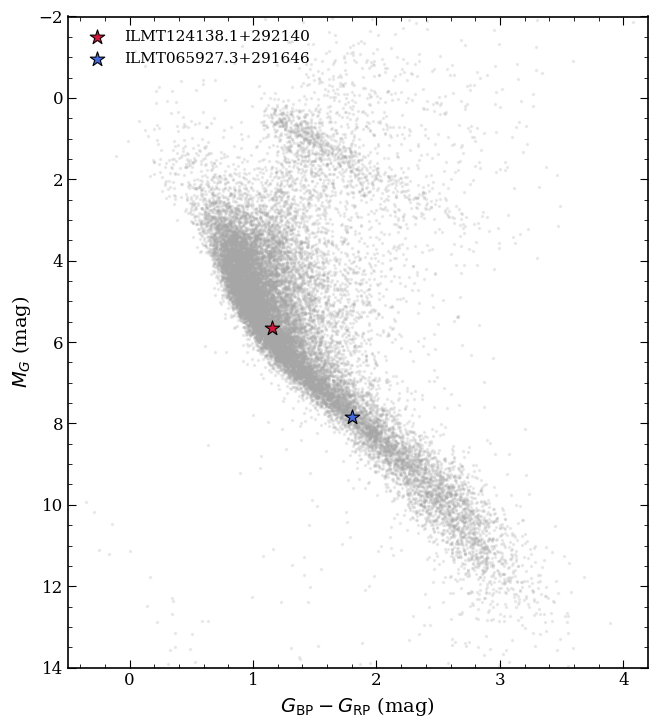}
    \caption{The two identified BY Draconis candidates highlighted in the Gaia color-magnitude diagram. Both candidates lie on the main sequence. The candidate ILMT124138.1+292140 is consistent with a K-type star, while ILMT065927.3+291646 is consistent with a late K to early M-type star.}
    \label{fig:Gaia_CMD}
\end{figure}

\begin{figure}[t]
    \centering

    \begin{minipage}{0.48\textwidth}
        \hspace*{-0.0cm}%
        \begin{minipage}{\linewidth}
            \centering
            \includegraphics[width=\linewidth]{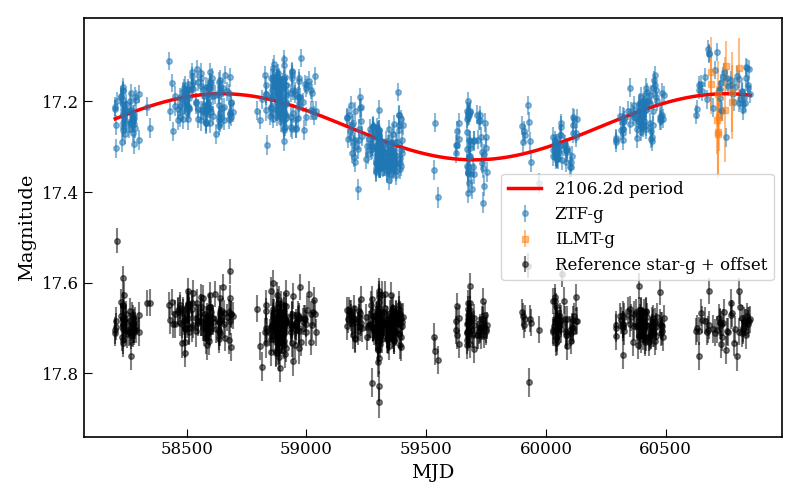}

            \hspace*{0.6cm}
            (a)
        \end{minipage}
    \end{minipage}


    \begin{minipage}{0.48\textwidth}
        \hspace*{-0.0cm}%
        \begin{minipage}{\linewidth}
            \centering
            \includegraphics[width=\linewidth]{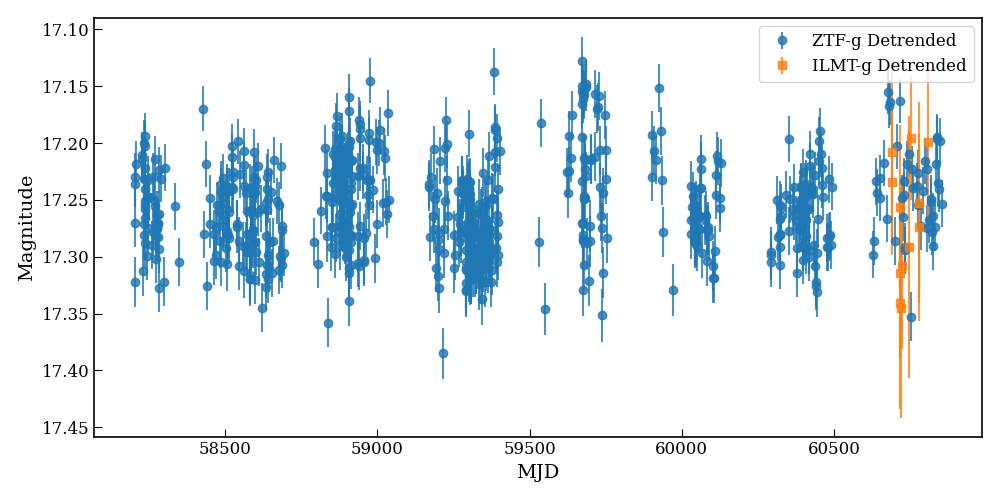}

            \hspace*{0.6cm}
            (b)
        \end{minipage}
    \end{minipage}


    \begin{minipage}{0.48\textwidth}
        \hspace*{-0.0cm}%
        \begin{minipage}{\linewidth}
            \centering
            \includegraphics[width=\linewidth]{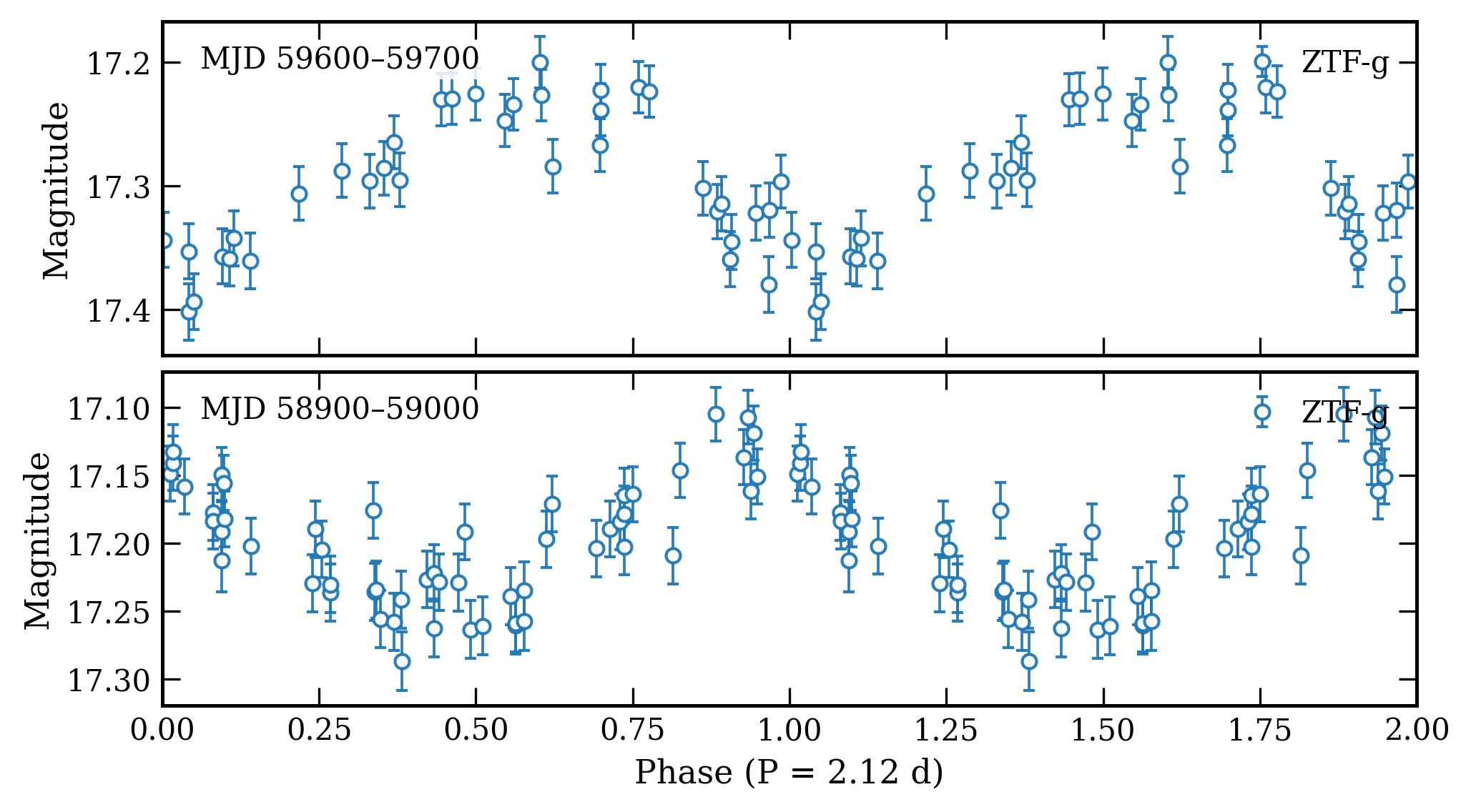}

            \hspace*{0.6cm}
            (c)
        \end{minipage}
    \end{minipage}

    \caption{Overview of variable star candidate ILMT124138.1+292140. (a) The raw g-band lightcurve shows clear long-term variability over $2106$ days. The lightcurve is also compared with a nearby non-variable reference star at $\mathrm{RA} = 12^{\mathrm{h}}41^{\mathrm{m}}48.20^{\mathrm{s}},\; \mathrm{Dec} = +29^\circ 24' 12.20''$ (b) The same g-band lightcurve detrended with a sinusoid. (c) Phase-folded detrended lightcurve, folded on a two different MJD bins (from MJD 58900 to 59000 and MJD 59600 to 59700) with a shorter period of $\sim$2.12 days. The two lightcurves indicate change in spot configuration from one MJD bin to another.}
    \label{fig:BY_draco_1}
\end{figure}

\begin{figure}[t]
    \centering

    \begin{minipage}{0.48\textwidth}
        \hspace*{-0.0cm}%
        \begin{minipage}{\linewidth}
            \centering
            \includegraphics[width=\linewidth]{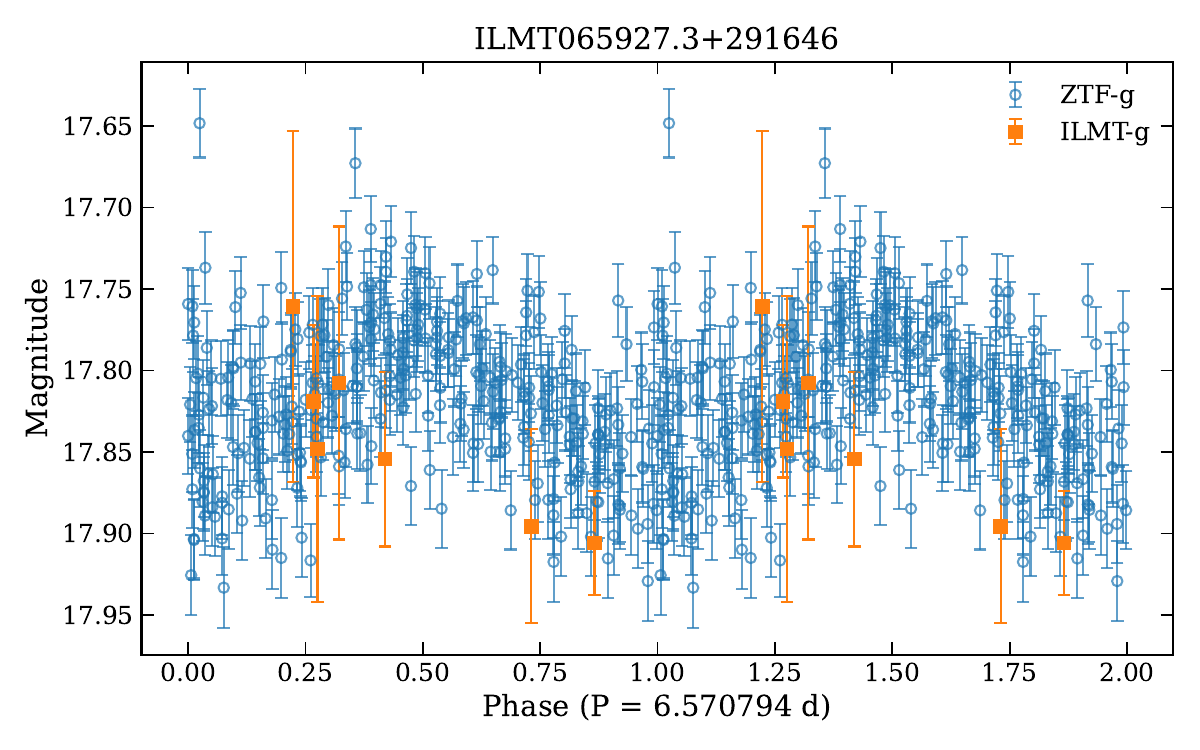}

            \hspace*{0.6cm}
            (a)
        \end{minipage}
    \end{minipage}


    \begin{minipage}{0.48\textwidth}
        \hspace*{-0.0cm}%
        \begin{minipage}{\linewidth}
            \centering
            \includegraphics[width=\linewidth]{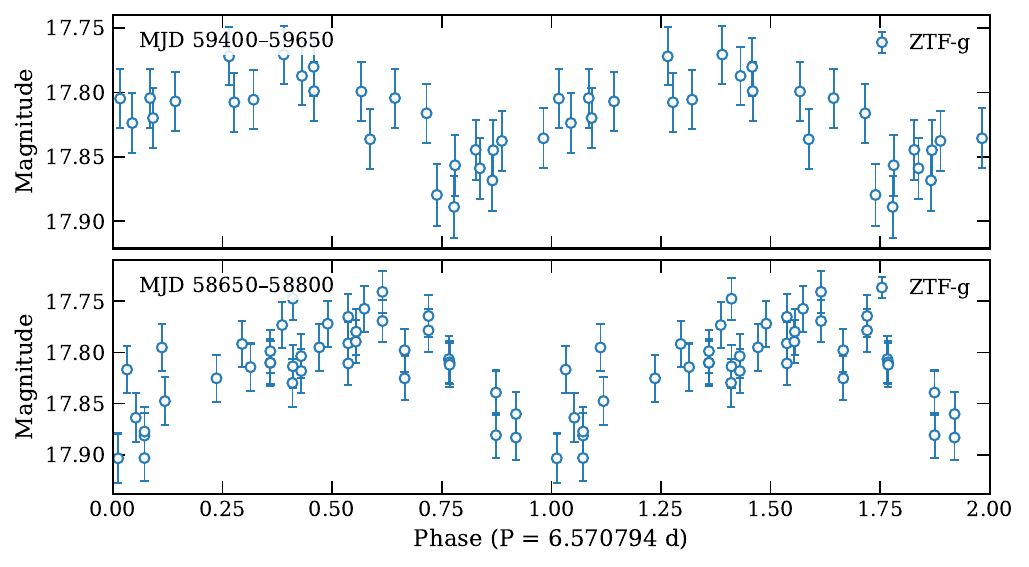}

            \hspace*{0.6cm}
            (b)
        \end{minipage}
    \end{minipage}

    \caption{Phase-folded lightcurves of the BY Draconis variable candidate ILMT065927.3+291646. (a) Combined phase-folded lightcurve using ZTF and ILMT g-band observations. (b) Phase-folded ZTF g-band lightcurves in two different MJD bins, showing evolution of the spot configuration.}
    \label{fig:BY_draco_2}
\end{figure}

\subsubsection{Identification of Two New BY Draconis Variable Candidates}
\label{sec:BY_Draconis}

During the search for variable sources which are uncatalogued in the VSX, two sources were identified that showed strong evidence of variability. These two sources could be immediately ruled out as eclipsing binary candidates, as they showed no evidence of a prominent eclipse-like feature. The sources lie clearly on the main sequence (Figure~\ref{fig:Gaia_CMD}), which further rules out the possibility of them being RR Lyrae candidates. Both sources are individually analyzed below, revealing that their variability is likely rotationally modulated by the presence of dark spots on their surfaces.

\subsubsection*{ILMT124138.1+292140}

ILMT124138.1+292140 ($\mathrm{RA} = 12^{\mathrm{h}}41^{\mathrm{m}}38.10^{\mathrm{s}},\; \mathrm{Dec} = +29^\circ 21' 40.0''$) is an unclassified variable source detected with the ILMT that shows a conspicuous long-term variability in its lightcurve as shown in the top panel (panel a) of Figure~\ref{fig:BY_draco_1}. The source lies on the main sequence and has a \textit{Gaia} $BP-RP$ color of $\sim$1.15, consistent with a K-type star \citep{2013ApJS..208....9P}. A potential periodicity of $\sim$2,106 days is evident in the long-term lightcurve; however, data spanning only a single cycle cannot conclusively confirm this trend. 
To search for short-term variability, the lightcurve was further detrended using a sinusoidal function with a period of $\sim$2,106 days, and the detrended lightcurve is shown in panel b of Figure~\ref{fig:BY_draco_1}.   
Furthermore, a periodogram analysis of the detrended lightcurve revealed a periodicity of $\sim$2.12 days. However, folding the complete lightcurve on this period resulted in a highly scattered phase-folded lightcurve. Consequently, the data were segmented into separate time intervals where continuous observations were available. The folded lightcurves of two time segments ( MJD 59600 to 59700 and from MJD 58900 to 59000)  of the detrended lightcurve are shown in panel c of Figure~\ref{fig:BY_draco_1}.
A secular variation is clearly evident in these lightcurves. Specifically, we observe changes in the shape, amplitude, and phase minimum across all segment-folded lightcurves. The evolution in the morphology of these phase-folded lightcurves indicates a changing spot configuration, indicating that the variability is driven by rotational modulation linked to BY Dra-type activity \citep[e.g.][]{2005AJ....130.1231P}. Considering the \textit{Gaia} Renormalized Unit Weight Error (RUWE) value for this source to be $\sim$1 (hence, favoring a single-star model) and it being a main-sequence star, most likely the variability in magnitude is modulated by star spots.
The long-term variability is most likely caused by stellar magnetic cycles with a period of $\sim$2106 days, which is expected for such a short rotation period active star \citep[see][]{2025MNRAS.540..668C}.

\subsubsection*{ILMT065927.3+291646}

ILMT065927.3+291646 ($\mathrm{RA} = 06^{\mathrm{h}}59^{\mathrm{m}}27.30^{\mathrm{s}},\; \mathrm{Dec} = +29^\circ 16' 46.0''$) is another uncatalogued main-sequence variable star, detected with the ILMT. The source lies on the main sequence and has a \textit{Gaia} $BP-RP$ color of $\sim$1.8, consistent with a late K to early M dwarf \citep{2013ApJS..208....9P}. A period of $\sim$6.57 days was determined for this star, which is consistent with the expected rotation periods for BY Dra-type variables. The \textit{Gaia} RUWE value for this star was determined to be $\sim$1, thus favoring a single star model and further supporting the source being a BY Draconis variable star. Figure \ref{fig:BY_draco_2}a displays the phase-folded lightcurve for the combined ZTF and ILMT $g$-band data, whereas Fig. \ref{fig:BY_draco_2}b illustrates the phase-folded lightcurves at two separate epochs. As observed in the source ILMT124138.1+292140, the variations in lightcurve morphology between epochs are consistent with spot-induced rotational modulation.

\subsubsection{Cataclysmic Variable Stars}
\label{sec:CV_outbursts}

\begin{figure}[ht]
    \centering
    \fbox{%
        \includegraphics[width=0.48\textwidth]{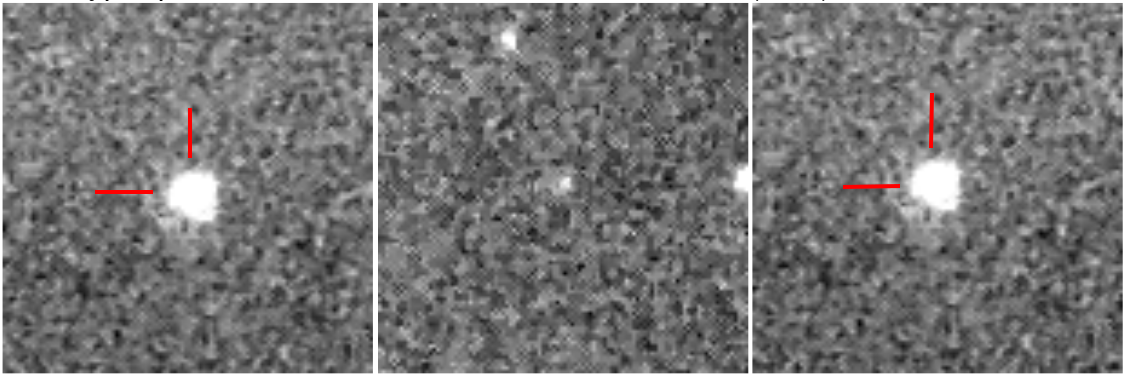}%
    }

    \vspace{0.4cm}

    \includegraphics[width=\linewidth]{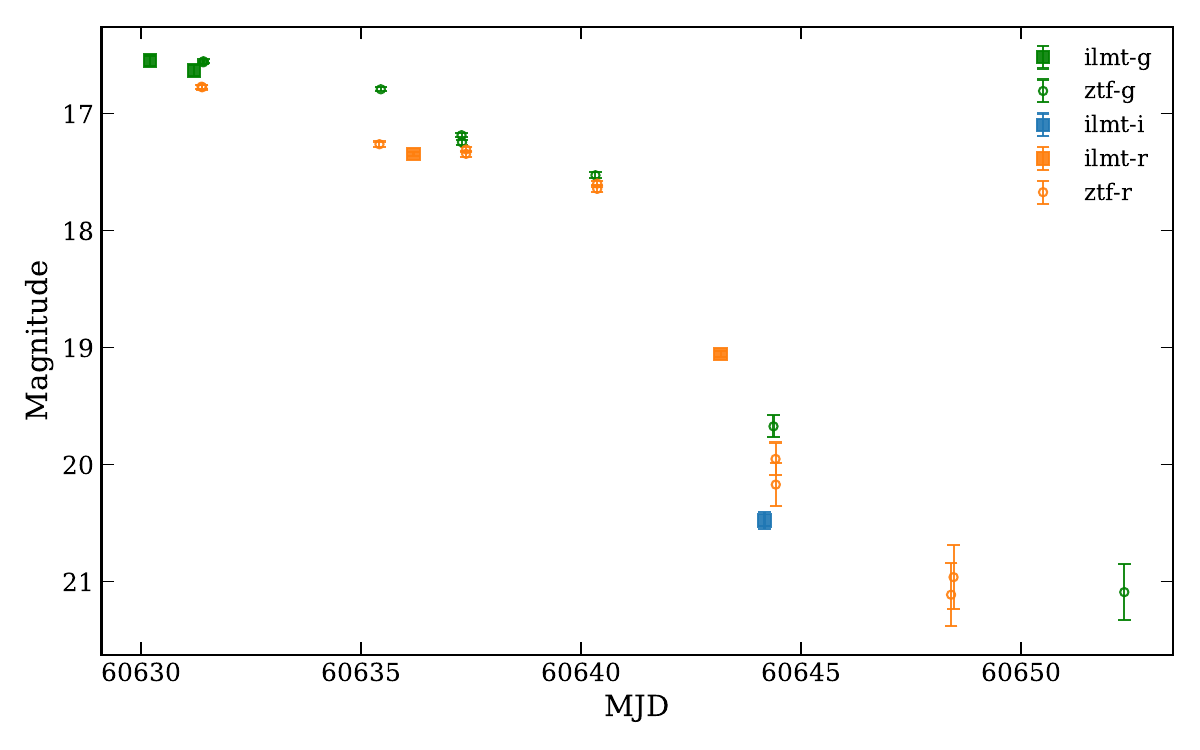}

    \caption{
    Top (left to right): Science, reference, and difference image cutouts of the cataclysmic variable candidate CRTS~J071924.1+292343, detected with the ILMT on 20~November~2024. Bottom: Multi-band light curve of the outburst event detected for the same source.}
    \label{fig:CRTS0719+2923}
\end{figure}

CV stars are binary systems consisting of a white dwarf and a secondary donor star, exhibiting episodes of mass transfer from the donor to the white dwarf. Such episodes are associated with the detection of outburst events in such systems. During outburst phases, such objects can become several magnitudes brighter, potentially enabling spectroscopic follow-up and detailed classification. The transient alert facility at the ILMT enables identification of such events and subsequent follow-up. Variability has been detected in 7 CV/CV--candidates with the ILMT, including outburst events detected in 3 unconfirmed candidates. Figure~\ref{fig:CRTS0719+2923} shows the lightcurve and detection images of one such outburst event associated with a CV candidate CRTS J071924.1+292343. The list of cataclysmic variables/cataclysmic variable candidates with detected variability and outburst episodes is shown in Table~\ref{tab:cataclysmic_vars}.  

\subsection{AGNs}

AGNs are galaxies that host supermassive black holes at their centres; they are actively accreting gas and emit radiation across a wide range of the electromagnetic spectrum, from X-rays to radio wavelengths. AGNs often exhibit flux variability and are routinely detected in optical surveys. Several sky surveys, such as the Sloan Digital Sky Survey \citep[SDSS;][]{2017ApJS..233...25A}, have aided in the discovery of hundreds of thousands of new AGNs. Modern time-domain surveys like the ZTF and the All Sky Automated Survey of SuperNovae \citep[ASSAS-SN;][]{2014AAS...22323603S} have enabled rich photometric datasets of such AGNs and have served as important diagnostic tools for understanding the variability and mechanisms behind the accretion process. To date, the ILMT has detected variability in 131 catalogued AGNs, including 105 quasars, 23 Seyfert 1 galaxies, and 3 blazars. Some of the interesting cases are discussed below.

\subsubsection{Search for highly variable quasars}

The application of difference imaging for transient detection in ILMT images inherently biases the search toward sources exhibiting large-amplitude photometric variability. While this selection limits sensitivity to low-level variability, it is particularly advantageous for identifying quasars undergoing significant brightness changes. Such highly variable quasars are of special interest for systematic searches of rare and scientifically compelling subclasses, including changing-look quasars (CLQs). To further investigate this population, we compiled a sample of quasars identified during the transient search that exhibit magnitude variations (magnitude difference between the brightest and faintest epochs) exceeding 1~mag in the ZTF \textit{r}$'$ or \textit{g}$'$ bands. The selected sources are presented in Table~\ref{tab:quasars}. An example light curve of a quasar showing large variability is shown in the left panel of Figure~\ref{fig:AGNs}.

\subsubsection{Optical support for blazar candidate NVSS J133101+293216}

\begin{figure*}
    \centering
    \begin{subfigure}[b]{0.47\textwidth}
        \centering
        \includegraphics[width=1.1\textwidth]{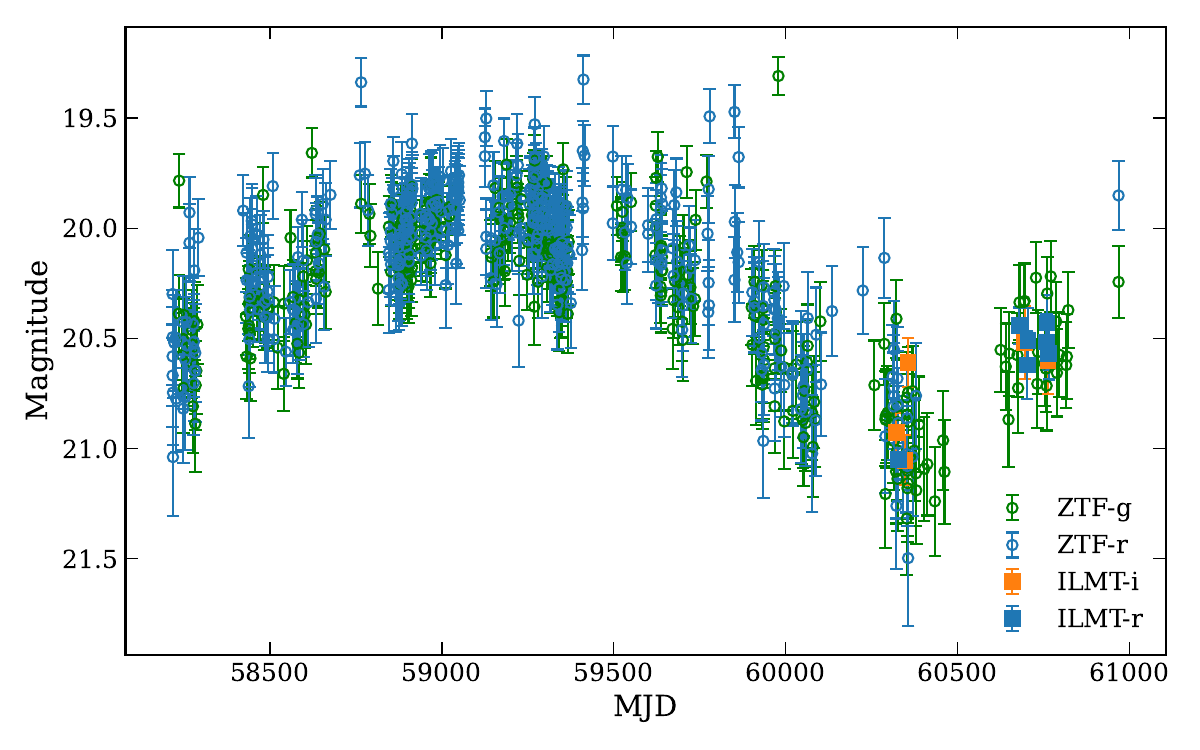}
        \label{fig:SDSS_J111253.99+293802.8}
    \end{subfigure}
    \hfill
    \begin{subfigure}[b]{0.47\textwidth}
        \centering
        \includegraphics[width=1.1\textwidth]{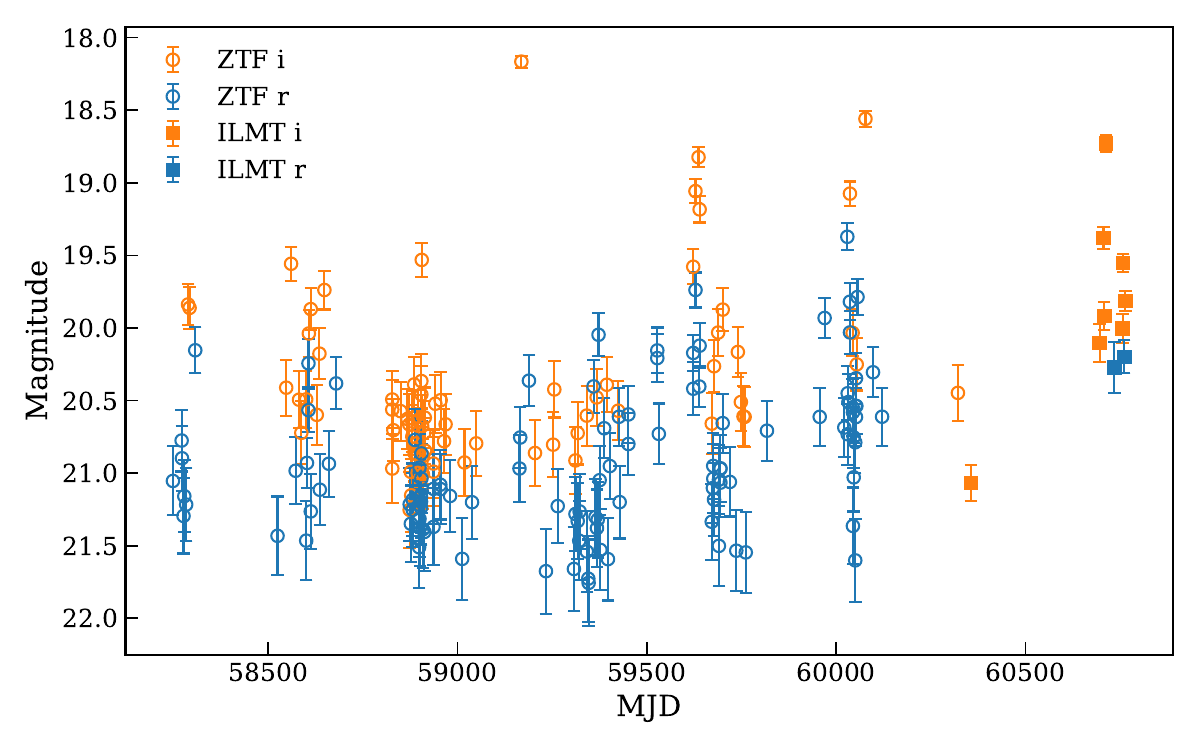}
        \label{fig:CRTS_CSS110428_293217_ir.pdf}
    \end{subfigure}
\vspace{-10pt}
\caption{Lightcurves of the quasar SDSS J111253.99+293802.8 showing variability by more than 1 mag in \textit{r}$'$ filter (left), and the blazar candidate NVSS J133101+293216 (right) detected with the ILMT.}
    \label{fig:AGNs}
\end{figure*}

NVSS J133101+293216 ($\mathrm{RA} = 13^{\mathrm{h}}31^{\mathrm{m}}01.84^{\mathrm{s}},\; \mathrm{Dec} = +29^\circ 32' 16.57''$) was identified as a blazar candidate in earlier studies by \citet{2018evn..confE.100G} and \citet{10.1093/mnras/stx396}, based on high-resolution radio imaging and $\gamma$-ray observations, respectively. The source was also identified as blazar candidate in recently released blazar and blazar candidates catalogues by \citet{2026A&A...708A.382K} and \citet{2026A&A...709A..37H}. Optical variability from this source was first detected with the ILMT on 7 February 2025. Following the ILMT detection, archival ZTF \textit{g}- and \textit{r}-band lightcurves were retrieved, as shown in Figure~\ref{fig:AGNs} (right panel). The source exhibits pronounced optical variability with an amplitude of approximately 2–2.5 magnitudes over a temporal baseline of $\sim$2000 days. This level of long-term optical variability is consistent with the blazar nature reported in earlier studies and provides additional optical support for the blazar interpretation. The candidate, when in flaring state, can be followed-up spectroscopically with the 3.6m DOT during subsequent follow-up observations. This can help in establishing the accurate blazar classification for this candidate. 

\subsection{Solar System Objects}

\begin{figure}
    \centering
    \includegraphics[width=0.48\textwidth]{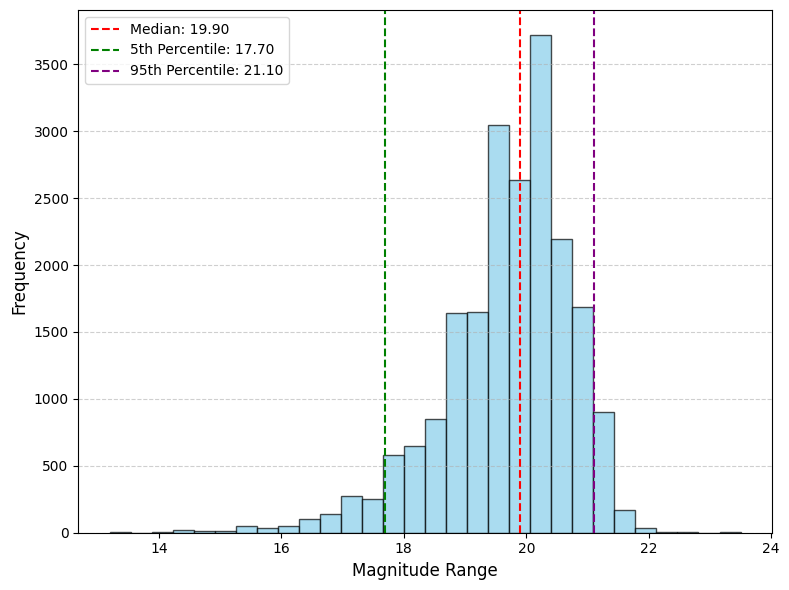}
    \caption{Distribution of MPC predicted V-band magnitudes of asteroids detected with the ILMT survey.}
    \label{fig:mag-asteroid}
\end{figure}

Solar system objects encompass comets, asteroids (including Trojans, Centaurs, and near-Earth objects (NEOs)), and trans-Neptunian objects. Asteroids, which appear as astrometric transients, are the most frequently detected class of objects in transient surveys like the ILMT. The faint limiting magnitude of approximately 22 in g$'$ band offers a distinct advantage by facilitating the detection of faint asteroids in a single exposure, since image stacking is not practical for non-sidereal objects such as asteroids. To identify asteroids, the SkyBoT\footnote{\url{https://ssp.imcce.fr/webservices/skybot/}} \citep{2006ASPC..351..367B} service of the IMCCE is queried corresponding to each transient candidate, with a search radius of 10$''$. From November 2023 to May 2025, more than 9000 different asteroids have been detected with the ILMT. In total, the pipeline recorded more than 20,000 asteroid detections, indicating multiple detections per object. Table~\ref{tab:asteroid_params} lists a few asteroids detected multiple times, along with their parameters. Figure~\ref{fig:mag-asteroid} illustrates the magnitude distribution of all the asteroid detections with the ILMT transient survey. 

\subsubsection{Near-Earth Objects}

Asteroids categorised as Near-Earth Objects (NEOs) have a perihelion distance $q < 1.3$ AU. This definition implies that their orbital paths bring them into the vicinity of Earth's orbit, making them dynamically distinct from main-belt asteroids and of particular interest due to their potential impact hazard and accessibility for spacecraft missions. To identify NEOs among the detected asteroids, the perihelion distances of all detected asteroids were obtained, yielding 19 NEOs. The orbital parameters of the detected NEOs, extracted from the NASA Small-Body Database Lookup\footnote{\url{https://ssd.jpl.nasa.gov/tools/sbdb_lookup.html}}, are shown in Table~\ref{tab:NEO_orbits}. Leveraging the high astrometric and photometric precision of the ILMT observations, the NEO observations can be used to more accurately constrain their orbital parameters.  

\begin{figure*}[ht]
    \centering
    \begin{minipage}[t]{0.48\textwidth}
        \centering
        \includegraphics[width=\linewidth]{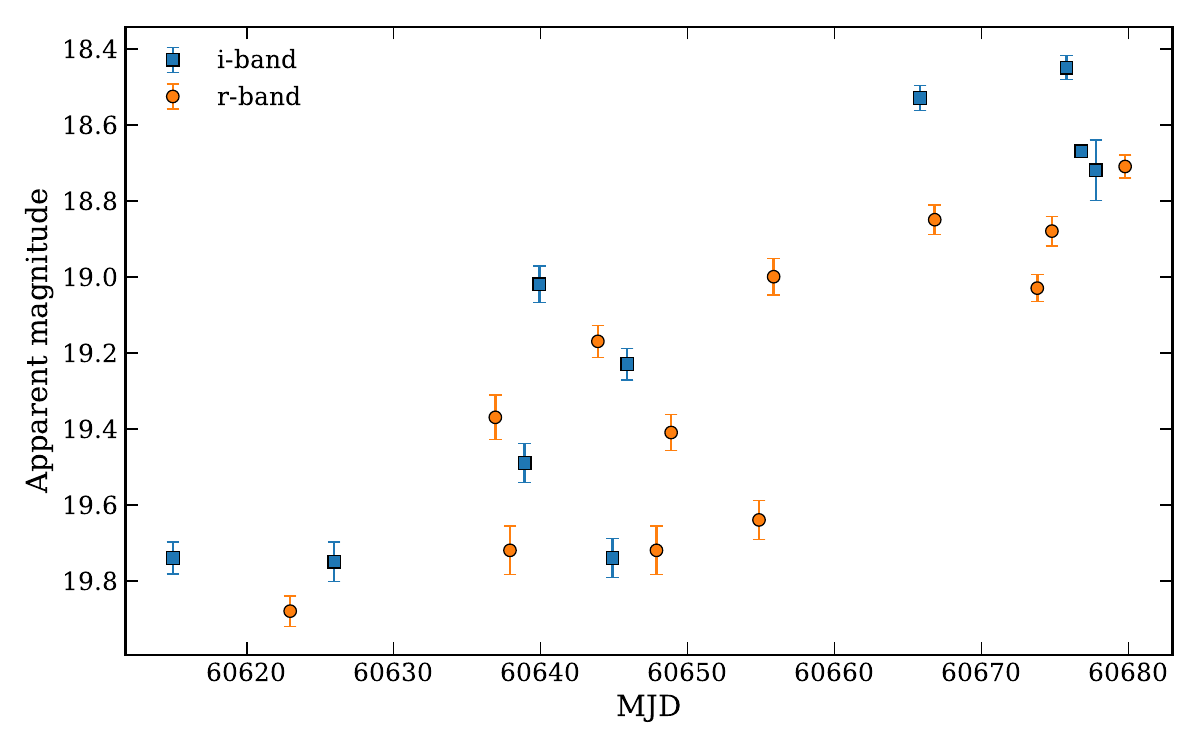}
    \end{minipage}
    \hfill
    \begin{minipage}[t]{0.48\textwidth}
        \centering
        \includegraphics[width=\linewidth]{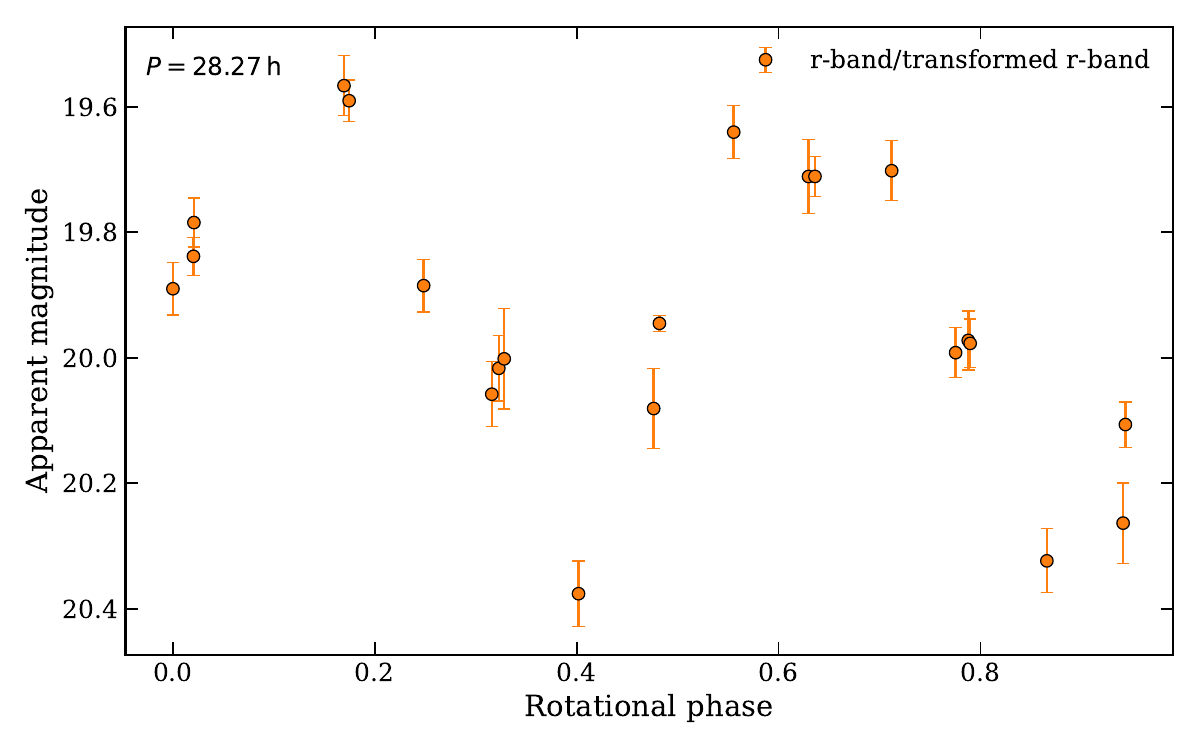}
    \end{minipage}
    
    \caption{Photometric observations of asteroid 2001~OW68. 
    The left panel shows the apparent magnitudes as obtained directly from observations, 
    while the right panel presents the orbital geometry–corrected and phase–folded light curve.}
    \label{fig:ow68_photometry}
\end{figure*}

\subsubsection{Asteroid 2001 OW68}

Due to the fixed-pointing nature of the ILMT, it is difficult to perform multi-epoch optical observations of asteroids, as they are moving objects. However, if the asteroid's trajectory is along the direction of image acquisition, multiple observations can be performed of a single asteroid on multiple nights. 2001 OW68 is one such asteroid that was observed on 27 nights with the ILMT. Figure \ref{fig:ow68_photometry} (left panel) illustrates the apparent magnitudes of 23 of the 27 detections for that asteroid in SDSS \textit{i}$'$ and \textit{r}$'$ filters. These magnitudes are modulated by the asteroid's geometric distances from the Earth and the Sun, solar phase angles, and asteroid rotation. The rotation-independent relation governing an asteroid's apparent magnitude can be modelled using the Equation \ref{eqn:ast_mag_equation} below.

\begin{equation}
\label{eqn:ast_mag_equation}
m = H + 5 \log \left( r \cdot \Delta \right) + 2.5 \log \left( \Phi(\alpha) \right)
\end{equation}

 where $H$ is the absolute magnitude of the asteroid, i.e., its magnitude at  1~AU from both the Sun and Earth at zero phase angle; $r$ is the heliocentric distance (in AU); $\Delta$ is the geocentric distance (in AU); $\alpha$ is the solar phase angle (in degrees); and $\Phi(\alpha)$ represents the phase function,  describing how the brightness of the asteroid changes with phase angle. For angles greater than $10^{\circ}$, a first-order approximation can be adopted,  where $2.5\log \left(\Phi(\alpha) \right) \sim \beta \cdot \alpha$, with $\beta$ being a constant linear coefficient expressed in magnitudes per degree. The effect of rotation can be obtained by correcting the observed magnitudes with modelled magnitudes. Figure \ref{fig:ow68_photometry} (right panel) illustrates phase-folded and corrected magnitudes. The period was determined using the string-length technique \citep{1983MNRAS.203..917D}, with the best-fitted r-i color index being 0.15. 

\section{\texttt{DART}: A \texttt{Streamlit}-Based Tool for Transient Alert Visualisation}
\label{sec:DART}

The \textbf{D}etection and \textbf{A}lert \textbf{R}eview \textbf{T}ool (\texttt{DART})\footnote{\url{https://ilmt-dart.shares.zrok.io/}} is an interactive web-based dashboard developed using the \texttt{Streamlit} framework to assist in the inspection and classification of transient candidates detected by the ILMT. The interface allows users to browse transient detections through a unified platform that combines tabular metadata with associated science, reference, and difference image cutouts.

The dashboard ingests transient candidate information from user-provided \texttt{CSV} files and consolidates it into a single data structure that includes relevant observational metadata, such as coordinates, timestamps, and host information. Basic categorisation is applied using \texttt{SIMBAD} classifications and flags for known solar system objects, enabling filtering by object type, such as variable stars, various AGN subtypes, and candidate extragalactic transients.

\begin{figure*}[ht]
    \centering
    \begin{minipage}{\textwidth}
        \centering
        
        \includegraphics[width=\linewidth]{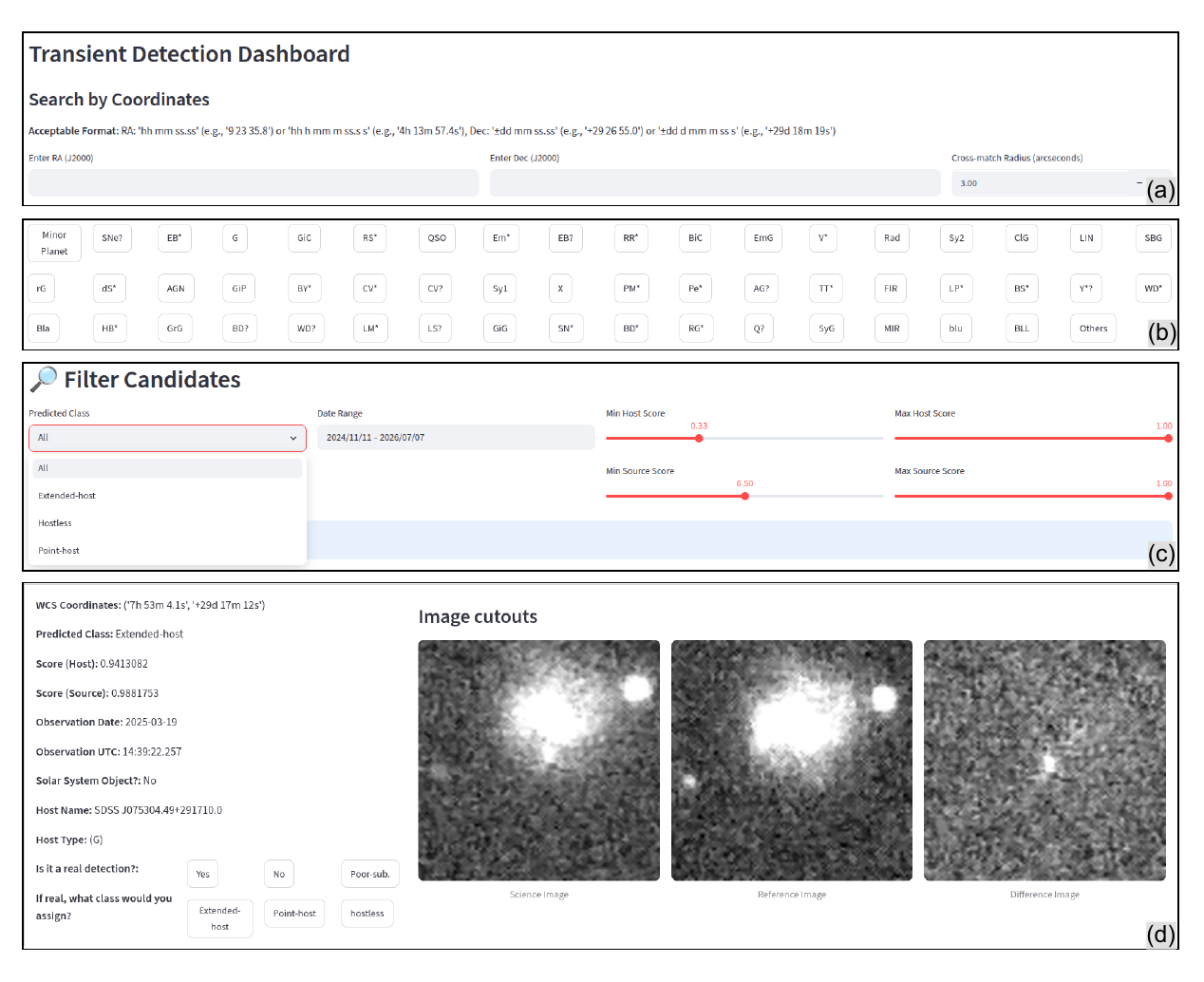}
        \caption{Various features available with \texttt{DART}, the transient alert dashboard developed for the ILMT. From top to bottom: (a) the cone-search interface for querying alerts by sky coordinates; (b) a button panel for filtering detections by their corresponding standard SIMBAD object types abbreviations\footnote{\url{https://simbad.cds.unistra.fr/Pages/guide/otypes.htx}}; (c) controls for filtering candidates based on the predicted transient class, detection date range, and ML confidence scores; and (d) the detailed view of an individual transient alert, displaying the science, reference, and difference image cutouts of a supernova candidate together with its key detection metadata. The interface also presents two survey questions that allow users to assess the validity of the detection and the predicted classification.}
        \label{fig:DART}
    \end{minipage}
\end{figure*}

Three primary exploration modes are supported. In the category-based view, users can filter detections by astrophysical class and inspect individual candidates via expandable panels that display observational details and image cutouts. A second mode enables coordinate-based searches, allowing users to query the dataset for detections within a specified radius of given sky coordinates. In the third mode, the user can filter candidates by host type and machine learning scores. Additionally, there is a provision for the user to provide feedback on detection and classification reliability. This will help to construct better datasets for training ML algorithms. Customisable features like these benefit from a custom-developed interface such as DART. Overall, \texttt{DART} provides a lightweight and efficient interface for reviewing transient alerts and validating detections within the ILMT survey pipeline. The GUI view of the framework is shown in Figure~\ref{fig:DART}.

\section{Discussion and Conclusion}
\label{sec:Discussion}

This work presents a census of transient and variable source detections obtained with ILMT from November 2023 to May 2025. The emphasis of this work is on detection, validation, and preliminary characterisation rather than on deriving full population statistics. The detected population spans a broad range of astrophysical classes, demonstrating the capability of ILMT to identify and characterise diverse transient and variable phenomena despite its limited FoV.

The 4-m ILMT achieved first light in April 2022 and has since been conducting a continuous astrometric and photometric survey of the zenith sky at a 1-day cadence. Using a dedicated transient detection and classification pipeline, \texttt{PyLMT}, the survey has been producing regular alerts for transient and variable sources. The detected candidates encompass a broad range of astrophysical classes, including SN candidates and confirmed SNe, AGNs such as QSOs, Seyfert galaxies, and blazars, as well as a diverse population of variable stars (including eclipsing binaries and RR Lyrae stars), cataclysmic variables, asteroids, and other transient and variable objects. The detected population also includes independent detections of newly identified SNe and variable star candidates. To facilitate public access to these alerts, an interactive web-based application, \texttt{DART}, was developed.

As the only operational liquid mirror telescope dedicated to time-domain astronomy, ILMT demonstrates that systematic transient searches can be conducted successfully using an unconventional telescope design. The stable observing geometry and uniform nightly cadence enable efficient difference imaging and variability detection, allowing scientifically robust transient searches within a narrow survey footprint. This makes ILMT a very cost-effective facility as compared to other conventional facilities having similar aperture size and depth. However, the limited FoV of the telescope should be taken into account.

Finally, this work illustrates the effectiveness of combining data from small, specialised surveys, such as ILMT, with those from wide-field facilities, such as ZTF. This synergy enables independent validation, extended temporal coverage, and improved characterisation of detected sources, reinforcing the continued scientific relevance of small surveys in the current era of time-domain astronomy. 

\section{Software and third party data repository citations} \label{sec:Software}

astropy\citep{2013A&A...558A..33A,2018AJ....156..123A,2022ApJ...935..167A}, astroquery \citep{2019AJ....157...98G}, photutils \citep{2016ascl.soft09011B}

\begin{acknowledgments}
We thank the referee for providing very constructive comments on the manuscript, which improved the clarity of the presentation. The 4-m International Liquid Mirror Telescope (ILMT) project results from a collaboration between the Institute of Astrophysics and Geophysics (the University of Li\`{e}ge, Belgium), the Universities of British Columbia, Laval, Montreal, Toronto, Victoria and York University, and Aryabhatta Research Institute of observational sciencES (ARIES, India). The authors thank Ankit Bisht, Hitesh Kumar, Himanshu Rawat, Khushal Singh, Nikhil Dharkiya, Rakesh Singh Bisht and other observing staff for their assistance at the ILMT.
KP acknowledges the support from the Erasmus+ Programme of the European Union for a research visit to the Institute of Astrophysics and Geophysics, University of Liège, Belgium (Allée du 6 Août 19c, 4000 Liège, Belgium). JS wishes to thank Service Public Wallonie, F.R.S.-FNRS (Belgium) and the University of Li\`{e}ge, Belgium, for funding the construction of the ILMT.  
PH acknowledges financial support from the Natural Sciences and Engineering Research Council of Canada, RGPIN-2019-04369. PH, AP-S, and JS thank ARIES for their hospitality during their visits to Devasthal. JS and KM acknowledge the assistance received from the Anusandhan National Research Foundation (ANRF, SERB- 762 VAJRA Faculty Scheme, India). KM, BA, and ND acknowledge support from the BRICS grant DST/ICD/BRICS/Call-5/CoNMuTraMO/2023 (G), funded by the DST, India.  
MD acknowledges the Innovation in Science Pursuit for Inspired Research (INSPIRE) fellowship award (DST/INSPIRE Fellowship/2020/IF200251). VN is supported by the Beijing Natural Science Foundation (Grant No. IS25004). 

\end{acknowledgments}

\begin{contribution}

All authors contributed equally to the ILMT collaboration and the resulting scientific outputs.


\end{contribution}

%
\facilities{4-m International Liquid Mirror Telescope (ILMT), 3.6m Devasthal Optical Telescope (DOT)}

\software{astropy \citep{2013A&A...558A..33A,2018AJ....156..123A,2022ApJ...935..167A}, astroquery \citep{2019AJ....157...98G}, photutils \citep{2016ascl.soft09011B}}


\begin{table*}[h]
\caption{Transient candidates detected with the ILMT and reported to the Transient Name Server (TNS). `ndet' refers to the number of automated detections for an event with the pipeline.}
\label{tab:ilmt_sne}
\begin{tabular}{c c c c c c c c}
\hline
Name & RA & Dec & Detection magnitude & Band (SDSS) & Remark & Type & ndet\\
\hline

AT 2023yjc & 01 50 02.900 & +29 08 53.00 & 20.6 & r & New discovery & - & 2\\
SN 2023vcg & 23 56 05.880 & +29 22 40.00 & 18.8 & r & - & IIP & 2\\
SN 2024cjb & 09 11 27.500 & +29 29 36.00 & 19.7 & r & - & Ia & 10\\
AT 2024ccg & 06 45 29.100 & +29 29 15.00 & 20.0 & i & - & - & 6\\
AT 2024eab & 08 24 49.500 & +29 36 49.00 & 19.4 & i & - & - & 6\\
AT 2024fxn & 14 13 49.600 & +29 23 32.00 & 19.7 & r & New discovery & - & 10\\
AT 2024fpx & 15 28 31.698 & +29 34 36.49 & 19.5 & r & - & - & 14\\
AT 2024zsm & 02 52 40.000 & +29 10 09.00 & 20.2 & r & New discovery & - & 6\\
AT 2024abso & 01 29 13.424 & +29 23 10.25 & 19.3 & i & - & - & 14\\
AT 2024agkc & 07 49 08.100 & +29 35 07.00 & 19.5 & r & New discovery & - & 4\\
AT 2025dip & 15 16 36.000 & +29 18 40.00 & 19.4 & r & - & - & 28\\
AT 2025dgl & 14 52 20.000 & +29 31 02.00 & 18.5 & r & - & - & 25\\
SN 2025dov & 11 16 00.000 & +29 21 48.00 & 18.0 & r & - & Ia & 20\\
AT 2025chp & 07 53 04.100 & +29 17 11.00 & 19.5 & r & - & - & 2\\
AT 2025re  & 13 31 15.600 & +29 22 16.00 & 19.4 & r & - & - & 30\\
AT 2024aifv & 16 12 11.100 & +29 34 17.00 & 20.3 & r & New discovery & - & 2\\
AT 2024aiha & 05 39 53.400 & +29 29 04.00 & 20.4 & r & New discovery & - & 8\\
\hline
\end{tabular}
\end{table*}

\begin{table*}[h]
\centering
\caption{Variable star candidates not present in the VSX and detected with the ILMT at least twice. All the candidates were cross-matched with the ZTF and had at least 10 detections. The variability parameter \textit{magsigma} was acquired from the ALeRCE broker of ZTF. The types of variable stars were determined from the lightcurves. The ILMT naming convention has been used as the source identifier and follows the coordinate-based \texttt{ILMTHHMMSS.S+DDMMSS} format.} 
\label{tab:variable_sources}
\begin{tabular}{cccccccc}
\hline
ILMT ID & ZTF OID & RA & Dec & magsigma\_g & magsigma\_r & PanSTARRS\_r & Type\footnote{EB: Eclipsing Binary, BY: BY-Draconis Variable, IRR: Irregular Variable}\\
 & & & & (mag) & (mag) & (mag) & \\
\hline
ILMT053553.5+292525 & ZTF18abmwwjd & 05 35 53.51 & +29 25 25 & 0.314 & 0.517 & 16.58 & EB\\
ILMT054850.6+292709 & ZTF18accckpk & 05 48 50.61 & +29 27 09 & -- & 0.132 & 19.36 & EB\\
ILMT055817.4+291600 & ZTF18abwcdxo & 05 58 17.44 & +29 16 00 & 0.355 & 0.323 & 16.99 & EB\\
ILMT063415.1+291359 & ZTF18abzccih & 06 34 15.14 & +29 13 59 & 0.202 & 0.198 & 17.99 & EB\\
ILMT065927.3+291646 & ZTF18aaesqpu & 06 59 27.33 & +29 16 46 & 0.144 & 0.206 & 16.70 & BY\\
ILMT070926.1+291835 & ZTF18aafahyp & 07 09 26.06 & +29 18 35 & 0.299 & 0.387 & 17.08 & EB\\
ILMT072913.0+293326 & ZTF19aarmugb & 07 29 13.02 & +29 33 26 & 0.112 & 0.424 & 17.96 & EB\\
ILMT080631.1+292601 & ZTF18aagacvx & 08 06 31.09 & +29 26 01 & -- & 0.211 & 17.58 & IRR\\
ILMT124138.1+292140 & ZTF19aagtvzm & 12 41 38.15 & +29 21 40 & 0.169 & 0.253 & 16.55 & BY\\
ILMT171652.1+292821 & ZTF18absjsmd & 17 16 52.06 & +29 28 21 & 0.275 & 0.237 & 16.42 & IRR\\
ILMT223112.7+291137 & ZTF18abnyfyb & 22 31 12.70 & +29 11 37 & 0.464 & 0.377 & 16.33 & EB\\
ILMT224904.4+290507 & ZTF18abilvtx & 22 49 04.44 & +29 05 07 & 0.376 & 0.425 & 17.25 & EB\\
ILMT065326.6+292147 & ZTF18aafdpnf & 06 53 26.63 & +29 21 47 & 0.211 & 0.164 & 17.95 & EB\\
\hline
\end{tabular}
\end{table*}

\begin{table*}[ht]
\centering
\caption{List of cataclysmic variables with detected variability.\footnote{Also includes a sample catalogued as an X-ray source but is also a CV candidate}}
\hspace*{-2cm}
\begin{tabular}{l c c c}
\hline
Name & RA & Dec & PanSTARRS\_r (mag) \\
\hline
CRTS J042430.6+292643 & 04 24 30.7 & +29 26 42 & 22.15\\
2CXO J044048.3+292434 & 04 40 48.3 & +29 24 34 & 20.00\\
CRTS J071924.1+292343 & 07 19 24.1 & +29 23 42 & --\\
1RXS J094558.3+292249  & 09 45 58.2 & +29 22 53 & 19.11\\
LAMOST J052658.99+291508.3  & 05 26 59.0 & +29 15 08 & 17.98\\
V$^{\ast}$ NY Her  & 17 52 52.5 & +29 22 19 & 18.40\\
TCP J06401599+2923447  & 06 40 16.0 & +29 33 45 & 22.70\\
\hline
\end{tabular}
\label{tab:cataclysmic_vars}
\end{table*}

\begin{table*}[ht]
\caption{Orbital parameters of selected asteroids detected multiple times with the ILMT.}
\begin{tabular}{lccccccc}
\hline
Asteroid & $a$ (AU) & $e$ & $i$ (°) & $q$ (AU) & $Q$ (AU) & $H$ (mag) & \textbf{ndet}\\
\hline
54892 (2001 OW68)       & 2.6791 & 0.1728 & 12.5328 & 2.2187 & 3.1394 & 14.73 & 27\\
143196 (2002 XE85)      & 2.6890 & 0.0782 & 7.1070  & 2.4786 & 2.8994 & 15.77 & 23\\
139868 (2001 RX69)      & 3.1497 & 0.1451 & 4.6153  & 2.6928 & 3.6066 & 15.14 & 23\\
248218 (2005 EA170)     & 3.0483 & 0.0537 & 11.4398 & 2.8847 & 3.2119 & 15.54 & 23\\
107813 (2001 FS59)      & 2.3399 & 0.1760 & 5.7289  & 1.9281 & 2.7517 & 16.00 & 22\\
88462 (2001 QM99)       & 2.4388 & 0.2303 & 5.5171  & 1.8772 & 3.0004 & 14.74 & 22\\
373802 (2002 VP41)      & 3.0924 & 0.1091 & 10.2874 & 2.7549 & 3.4298 & 15.57 & 21\\
56706 (2000 LD36)       & 2.4131 & 0.0200 & 7.1161  & 2.3649 & 2.4613 & 15.46 & 21\\
71840 (2000 US74)       & 2.2539 & 0.1238 & 5.2185  & 1.9749 & 2.5328 & 15.83 & 20\\
79241 Fulviobressan     & 2.3787 & 0.1326 & 4.0055  & 2.0634 & 2.6940 & 16.35 & 20\\
\hline
\end{tabular}
\label{tab:asteroid_params}
\end{table*}

\begin{table*}[h!]
\caption{Detected Near-Earth Objects (NEOs) with the ILMT and their orbital parameters from NASA JPL Small-Body Database Lookup.}
\label{tab:NEO_orbits}
\begin{tabular}{lccccccc}
\hline
\textbf{Name} & \textbf{$a$ (AU)} & \textbf{$e$} & \textbf{$i$ ($^\circ$)} & \textbf{$q$ (AU)} & \textbf{$ad$ (AU)} & \textbf{$H$ (mag)} \\
\hline
887 Alinda (A918 AA) & 2.473 & 0.5712 & 9.4 & 1.061 & 3.89 & 13.81 \\
25916 (2001 CP44) & 2.558 & 0.4995 & 15.76 & 1.28 & 3.84 & 13.77 \\
40267 (1999 GJ4) & 1.339 & 0.8084 & 34.48 & 0.257 & 2.42 & 15.54 \\
137925 (2000 BJ19) & 1.292 & 0.7638 & 31.12 & 0.305 & 2.28 & 16.07 \\
138893 (2000 YH66) & 1.173 & 0.7437 & 18.36 & 0.301 & 2.05 & 18.19 \\
144861 (2004 LA12) & 2.509 & 0.7491 & 39.4 & 0.63 & 4.39 & 15.32 \\
155334 (2006 DZ169) & 2.035 & 0.4087 & 6.61 & 1.204 & 2.87 & 17.23 \\
162004 (1991 VE) & 0.8908 & 0.6647 & 7.22 & 0.299 & 1.48 & 18.29 \\
185851 (2000 DP107) & 1.365 & 0.3766 & 8.67 & 0.851 & 1.88 & 18.31 \\
226514 (2003 UX34) & 1.095 & 0.6158 & 2.56 & 0.421 & 1.77 & 20.14 \\
363267 (2002 GS) & 1.353 & 0.4038 & 19.76 & 0.807 & 1.9 & 20.05 \\
380981 (2006 SU131) & 1.728 & 0.3953 & 9.66 & 1.045 & 2.41 & 18.64 \\
420187 (2011 GA55) & 2.169 & 0.4832 & 8.72 & 1.121 & 3.22 & 18.16 \\
499490 (2010 MW) & 2.277 & 0.4711 & 21.35 & 1.205 & 3.35 & 18.98 \\
613986 (2008 JG) & 1.051 & 0.2958 & 7.92 & 0.74 & 1.36 & 20.81 \\
762379 (2011 CG2) & 1.177 & 0.1585 & 2.76 & 0.991 & 1.36 & 21.41 \\
800056 (2014 AB55) & 2.369 & 0.485 & 16.57 & 1.22 & 3.52 & 18.47 \\
(2001 QB34) & 2.206 & 0.4175 & 5.74 & 1.285 & 3.13 & 19.84 \\
(2010 XY82) & 2.188 & 0.4896 & 26.78 & 1.117 & 3.26 & 19.17 \\
\hline
\end{tabular}
\end{table*}

\afterpage{\clearpage\begin{longtable*}{lrrrr}
\caption{List of quasars detected by the ILMT that show variability greater than 1 mag in archival ZTF \textit{g}$'$ or \textit{r}$'$ observations.} 
\label{tab:quasars} \\
\toprule
NAME & RA & Dec & Amp\_r (mag) & Amp\_g (mag) \\
\midrule
\endfirsthead

\caption[]{List of quasars detected by the ILMT (continued)} \\
\toprule
NAME & RA & Dec & Amp\_r (mag) & Amp\_g (mag) \\
\midrule
\endhead

\midrule
\multicolumn{5}{r}{\textit{Continued on next page}} \\
\endfoot

\bottomrule
\endlastfoot

LAMOST J143052.42+293633.9  & 14 30 52.4 & +29 36 34 &  1.10  & 1.21 \\
SDSS J111031.83+292738.0  & 11 10 31.8 & +29 27 38 &  1.52 & 1.01 \\
LAMOST J164048.01+293446.7  & 16 40 48.0 & +29 34 47 &  1.14  & 1.08 \\
SDSS J132209.99+293447.1  & 13 22 10.0 & +29 34 47 &  1.13  & 1.20 \\
LAMOST J114903.39+294005.3  & 11 49 03.4 & +29 40 05 &  1.71  & 1.20 \\
SDSS J153838.93+291803.6  & 15 38 39.0 & +29 18 05 &  1.24  & 1.22 \\
SDSS J123447.21+292434.5  & 12 34 47.3 & +29 24 35 &  1.21  & 0.91 \\
SDSS J104256.60+293938.2  & 10 42 56.6 & +29 39 39 &  1.14  & 1.15 \\
SDSS J103726.98+292052.7  & 10 37 27.0 & +29 20 54 &  1.49  & 1.17 \\
SDSS J123642.76+292449.4  & 12 36 42.8 & +29 24 50 &  1.08  & 1.12 \\
SDSS J145253.91+293224.1  & 14 52 54.0 & +29 32 24 &  1.61  & 1.44 \\
LAMOST J140917.10+292025.8  & 14 09 17.1 & +29 20 26 &  1.72  & 1.26 \\
US 2988  & 11 48 43.4 & +29 29 35 &  1.24  & 1.04 \\
SDSS J104323.05+292310.9  & 10 43 23.1 & +29 23 11 &  1.23  & 0.99 \\
SDSS J140500.81+292514.0  & 14 05 00.8 & +29 25 15 &  1.35  & 1.62 \\
2MASS J09113503+2938038  & 09 11 35.0 & +29 38 04 &  0.83  & 1.24 \\
SDSS J094319.33+293900.4  & 09 43 19.3 & +29 39 01 &  1.06  & 2.10 \\
2MASS J09465955+2932516  & 09 46 59.6 & +29 32 51 &  1.04  & 0.86 \\
US 3097  & 11 52 58.7 & +29 30 14 &  1.04  & 1.10 \\
SDSS J095943.84+292506.3  & 09 59 43.8 & +29 25 07 &  1.02  & 1.03 \\
SDSS J083422.51+292018.3  & 08 34 22.5 & +29 20 18 &  1.13  & 1.01 \\
SDSS J111253.99+293802.8  & 11 12 54.0 & +29 38 03 &  2.01  & 2.17 \\
VV2006 J081027.9+293045  & 08 10 27.9 & +29 30 45 &  1.20  & 1.10 \\
VV2006 J080651.7+293024  & 08 06 51.7 & +29 30 25 &  2.45  & 1.42 \\
SDSS J104643.63+292733.8  & 10 46 43.6 & +29 27 34 &  0.77  & 1.04 \\
SDSS J100836.89+292252.6  & 10 08 36.9 & +29 22 53 &  1.17  & 1.33 \\
LAMOST J121055.24+293718.1  & 12 10 55.3 & +29 37 18 &  1.45  & 1.45 \\
SDSS J113003.60+291933.1  & 11 30 03.7 & +29 19 34 &  1.34  & 1.37 \\
SDSS J115127.12+293743.8  & 11 51 27.2 & +29 37 44 &  1.53  & 1.83 \\
SDSS J122808.19+293138.8  & 12 28 08.2 & +29 31 38 &  1.08  & 0.86 \\
SDSS J115320.23+292226.7  & 11 53 20.3 & +29 22 26 &  1.79  & 1.34 \\
SDSS J100728.01+293731.8  & 10 07 28.0 & +29 37 32 &  1.19  & 1.24 \\
SDSS J104145.02+293200.5  & 10 41 45.0 & +29 32 01 &  1.30  & 1.44 \\
SDSS J235322.70+290651.6  & 23 53 22.7 & +29 06 53 &  1.14  & 1.03 \\
SDSS J101845.40+293741.6  & 10 18 45.4 & +29 37 43 &  1.45  & 1.63 \\
SDSS J100541.21+292520.8  & 10 05 41.2 & +29 25 21 &  1.11  & 1.23 \\
SDSS J092728.44+293503.6  & 09 27 28.5 & +29 35 04 &  0.72  & 1.01 \\
SDSS J093042.26+293419.2  & 09 30 42.3 & +29 34 19 &  1.04  & 1.17 \\
QSO J0937+2937  & 09 37 04.0 & +29 37 06 &  1.20  & 0.83 \\
SDSS J021147.03+292152.0  & 02 11 47.0 & +29 21 52 &  1.05  & 1.52 \\
SDSS J085200.66+291903.8  & 08 52 00.7 & +29 19 04 &  1.79  & 1.52 \\
SDSS J082926.57+292901.2  & 08 29 26.6 & +29 29 01 &  0.96  & 1.26 \\
VV2006 J081514.9+293547  & 08 15 14.9 & +29 35 49 &  1.39  & 0.98 \\
2MASS J07394488+2933122  & 07 39 45.0 & +29 33 13 &  1.18  & 1.02 \\
SDSS J014550.76+292305.0  & 01 45 50.8 & +29 23 06 &  1.02  & 1.03 \\
VV2006 J075535.4+292047  & 07 55 35.4 & +29 20 47 &  1.19  & 1.22 \\
SDSS J220938.09+291752.7  & 22 09 38.0 & +29 17 53 &  1.90  & 1.87 \\
SDSS J222329.31+291114.7  & 22 23 29.3 & +29 11 15 &  0.94  & 1.02 \\
SDSS J015831.15+292024.1  & 01 58 31.2 & +29 20 24 &  1.75  & 1.43 \\
SDSS J014749.39+291610.2  & 01 47 49.4 & +29 16 10 &  1.82  & 2.60 \\
SDSS J015539.87+290755.7  & 01 55 39.8 & +29 07 56 &  0.91  & 1.20 \\
SDSS J114740.38+292109.6  & 11 47 40.4 & +29 21 10 &  1.27  & 1.36 \\
SDSS J102712.39+293728.9  & 10 27 12.4 & +29 37 29 &  2.41  & 1.51 \\
SDSS J221646.70+292050.8  & 22 16 46.7 & +29 20 51 &  1.51  & 1.50 \\
SDSS J222343.72+290617.8  & 22 23 43.7 & +29 06 17 &  1.11  & 1.22 \\
\hline

\end{longtable*}}
\clearpage




\bibliography{ILMT_transient_survey}{}
\bibliographystyle{aasjournalv7}



\end{document}